\documentclass[%
 reprint,
 amsmath,amssymb,
 aps,
prb,
]{revtex4-1}

\usepackage{graphicx}
\usepackage{dcolumn}
\usepackage{bm}
\usepackage{subcaption}

\begin{document}

\preprint{APS/123-QED}

\title{Strain-based Spin Manipulation on Substitutional Nickel in Silicon Carbide}

\author{Wenhao Hu}
\author{Michael E. Flatt\'{e}}%
\affiliation{%
 Department of Physics and Astronomy, The University of Iowa, Iowa City, Iowa 52242-1479 USA\\
}%




\date{\today}

\begin{abstract}
By using the full potential linear augmented plane wave (FP-LAPW) method and full potential local orbital minimum basis (FP-LOMB) method within generalized gradient approximation (GGA), we studied the electronic structures and magnetic properties of nickel and chromium single dopants in polytypes of silicon carbide (SiC). The magnetic phases of defects are found to be strongly dependent on the external stress on the supercell. In 3C-SiC, the Ni single dopant exhibits an anti-ferromagnetic (AFM) to ferromagnetic (FM) transition at a moderate compressive and tensile hydrostatic strain in Si-sub and C-sub cases. In contrast, the Ni single dopant in 4H-SiC is stably in the nonmagnetic phase under external stress. The Cr single dopant is also insensitive to the applied stress but stably in the magnetic phase. This strain controlled magnetic transition makes the Ni single dopant a novel scheme of qubit.
\begin{description}
\item[PACS numbers]
\end{description}
\end{abstract}

\pacs{Valid PACS appear here}
\maketitle


\section{\label{sec:level1}Introduction}

In the synthesis of diamond by high-pressure and high-temperature technique, transition metals are important catalysts to accelerate the conversion from graphite to diamond, during which the nickel complexes are incorporated into the diamond matrix in different forms. Through the demonstration of electron paramagnetic resonance, photoluminesence and optical absorption, numerous nickel-related defects has been identified in the diamond, including substitutional nickel\cite{Isoya1990}, nickel-vacancy\cite{Iakoubovskii2004} and nickel-vacancy-nitrogen complexes (NE centers)\cite{Nadolinny1999}. Several theoretical approaches have been used to investigate these nickel-related defects' structural and electronic properties\cite{Gerstmann2000,Johnston2003,Pereira2003}. Experimentally, it has been reported that NE8 center can be an excellent infrared single-photon source at room temperature\cite{Gaebel2004,Rabeau2005,Wu2007}. 

In spite of that the NV center in diamond is an ideal candidate of qubit, it is still difficult to fabricate devices from diamond. In contrast, SiC is a wide-bandgap semiconductor with mature growth and device engineering technique. Commercial epitaxial and bulk monocrystal SiC are both available with high quality\cite{POWELL2006}. Intuitively, the single defect in SiC should have a long coherence time since the host atoms have stable spinless nuclear isotopes. In addition, the wide bandgap of SiC enables it to host various color centers.


\begin{figure}[h!]
  \centering
    \includegraphics[width=0.45\textwidth]{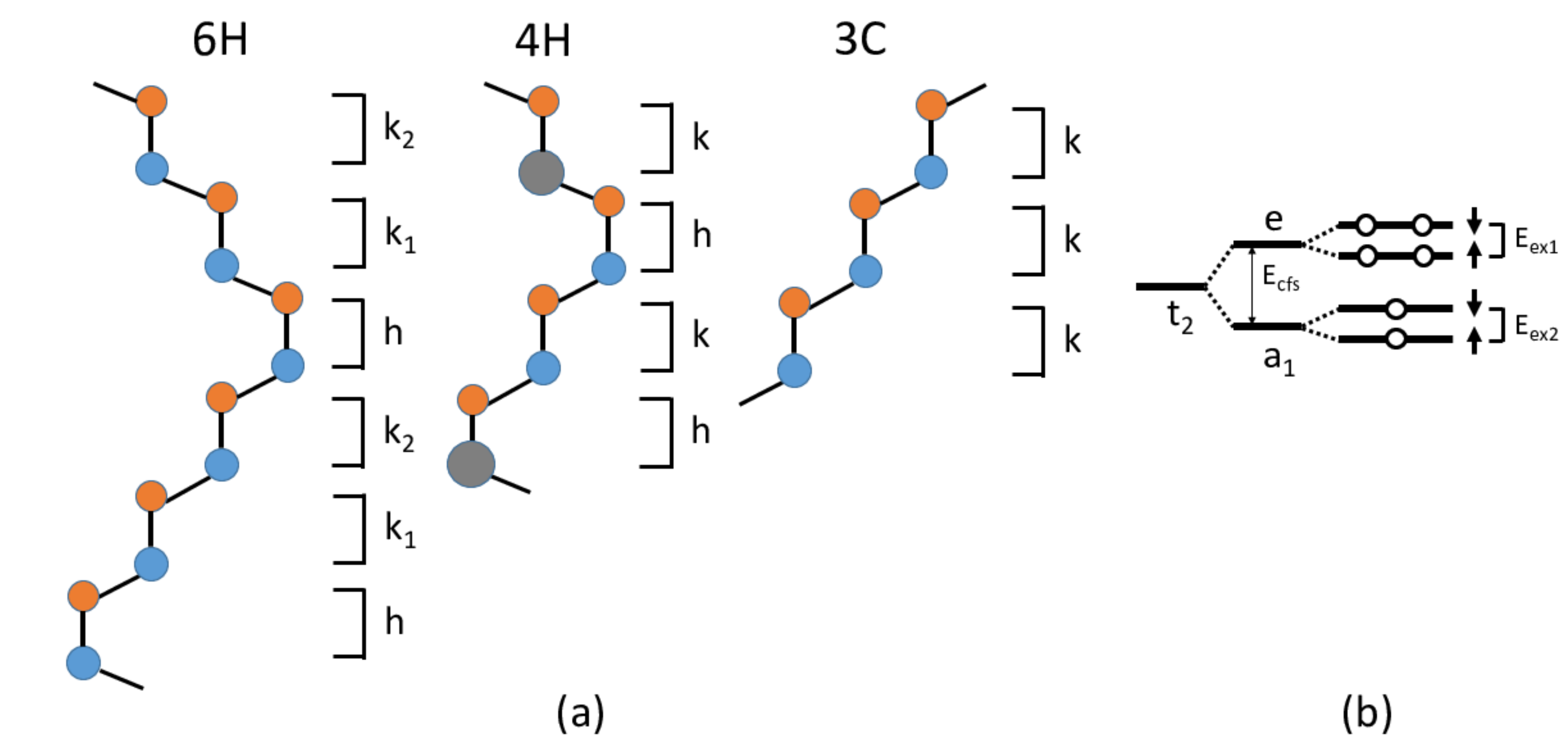}
  \caption{(a) Sequence of SiC stack. (b) Crystal field splitting due to symmetry lowering.}
  \label{SiCstack}
\end{figure}

But different from diamond, silicon carbide has various polytypes based on different stacking sequences of Si-C bilayer. More specificly, the atomic configuration of silicon carbide can be regarded as an alternating stack of two-dimensional Si and C layers. As shown in Fig.\ref{SiCstack}, two different Si-C bilayers (h and k) are ordered in different patterns to produce 3C, 4H and 6H-SiC. Note that there will be many an inequivalent substitutional sites for both Si and C in hexagonal SiC. Thousands of structures of SiC have been discovered obeying above rules, of which 3C, 4H and 6H SiC are three of the most common polytypes. They have pervasive appication in the power, opto-electronics and are also used as substrate of graphene\cite{Berger2006} and gallium nitride\cite{Liu2002}. Combined with magnetic resonance measurement and numerical simulation, it is predicted recently that several defects in SiC is also possible to be competitive candidates as qubit\cite{Mizuochi2002,Son2003,Son2006,GaliBSP2011}. For instance, it is recently demonstrated that there exists optical addressable spin states with long coherence times (5-50 $\mu s$) at room temperature\cite{Falk2013}. Double electron-electron resonance measurement in 6H-SiC proves the dipole-dipole coupling between spin ensembles, which is significant in practical spin-based quantum technologies. In this paper, we will investigate the possibility of realize universal manipulation of spin state of nickel dopant in polytypes of SiC with strain and microwave radiation.

This  paper is organized as follows: In the Sec. \ref{compdetails}, we summarized the parameters and methods of our calculations in this paper and the relaxed geometries of nickel complexes. In addition, the Si-C charge transfer in SiC is also shown, which is a significant difference from diamond. Then, magnetic and electronic properties of substitutional nickel (Ni$_\textrm{s}$) in silicon site and carbon site under hydrostatic strain will be discussed. 3C (Sec. \ref{SiC_c} and \ref{SiC_si}) and 4H (Sec. \ref{4HSiC}) polytypes of silicon carbide are both considered. Next, we will talk about the implementation scheme of Ni$_\textrm{s}$ in SiC and its prospective directions to dig. In Sec. \ref{Cr_SiC}, we want to briefly talk about the electronic structure of chromium single dopant in 3C/4H-SiC. In the end, we will summarize our results and make a general comment in Sec. \ref{conclusion}.

\section{Computational details}
\label{compdetails}

\begin{figure}
    \begin{subfigure}[b]{0.28\textwidth}
        \includegraphics[width=\textwidth]{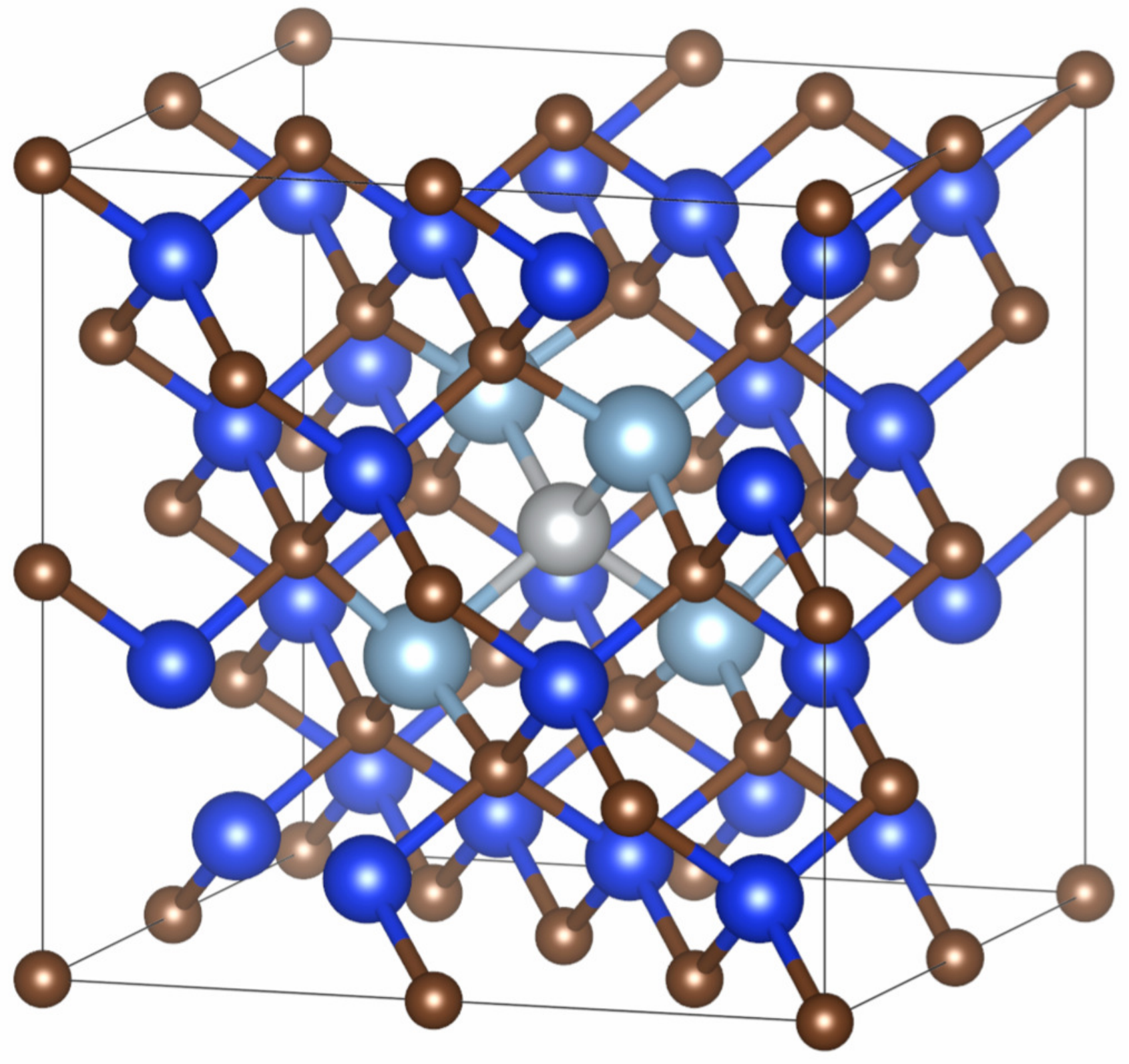}
        \caption{}
        \label{fig:3C}
    \end{subfigure}
    ~ 
    \begin{subfigure}[b]{0.17\textwidth}
        \includegraphics[width=\textwidth]{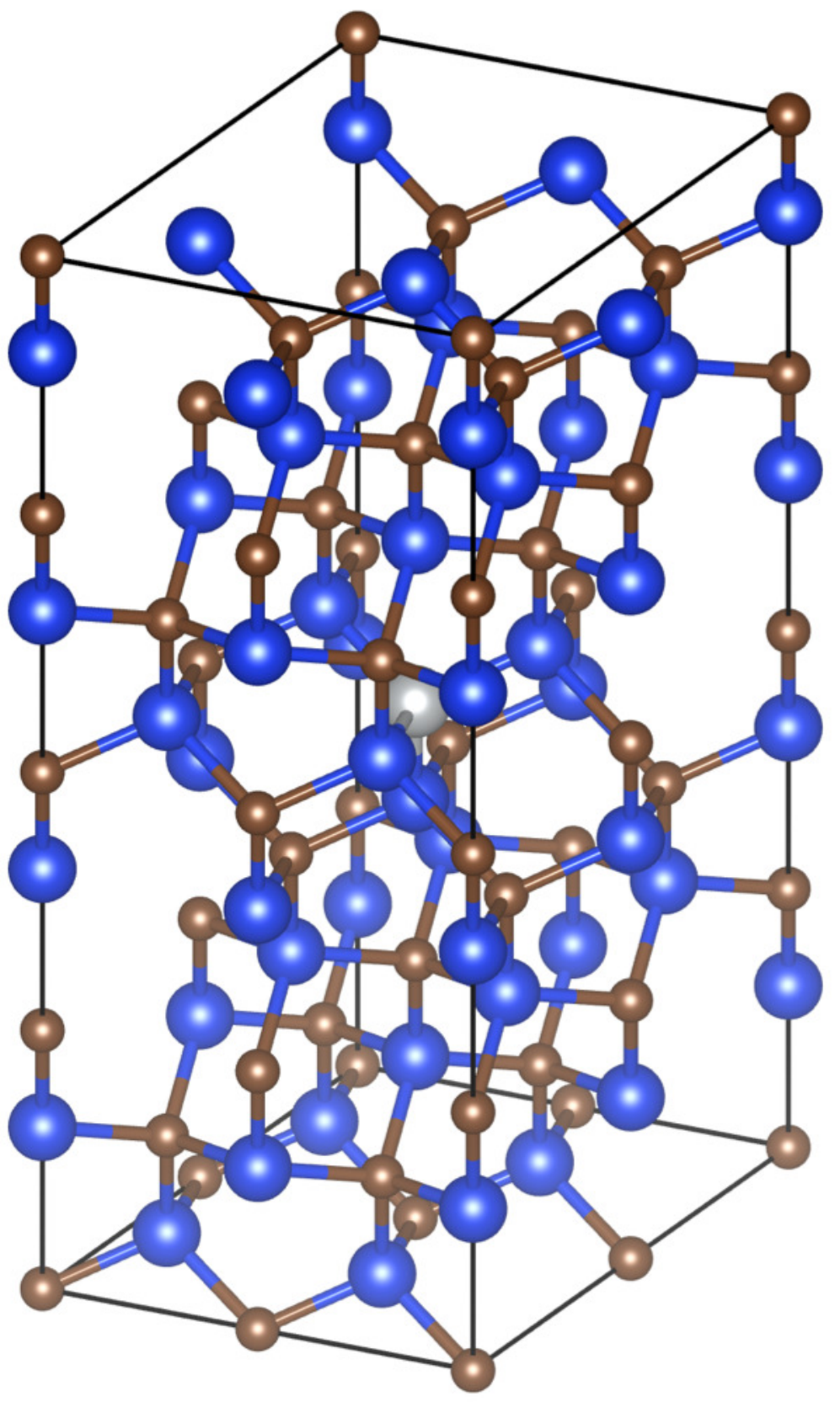}
        \caption{}
        \label{fig:4H}
    \end{subfigure}
    \caption{Nickel doped (a) 3C and (b) 4H SiC supercells.}
    \label{SiCsupercell}
\end{figure}

In our calculation, the FP-LAPW method implemented in WIEN2K package is exploited\cite{Schwarz2002}. The PBE-GGA\cite{PBE1996} is used for the exchange correlation functional in the DFT self-consistent calculations. To achieve an accuracy of 1 meV in total energy, a $7\times7\times7$ Monkhorst-Pack k-mesh\cite{Monkhorst1976} is exploited. The cutoff energy is set up according to the product of minimum radius of muffin-tin and maximum wave vector $RKmax=7.0$. Due to the periodic boundary condition used in the calculation, the interaction between neighboring unit cell need to eliminated through the method of supercell. Considering the tradeoff between the limit of computational resource and accuracy, a $2\times2\times2$ supercell is considered. As shown in Fig.\ref{SiCsupercell}, the impurity-contained supercells are NiSi$_\textrm{x}$C$_\textrm{63-x}$($x=31$ or $32$) supercells that consist of $2\times2\times2$ multiple of 3C (Fig.\ref{fig:3C}) and 4H (Fig.\ref{fig:4H}) SiC unit cells with a central nickels. The supercell size is determined according to experimental lattice constants of SiC: $a_{3c}=4.3596\ \AA$, $a_{4h}=3.0730\ \AA$ and $b_{4h}=10.053\ \AA$. All the atoms are allowed to relax with a precision of 1meV/$\AA$ according to T$_d$ symmetry. The spin-orbital interaction is not included. In this paper, only hydrostatic strain is considered, which can be realized by proportionally extending or shrinking the volume of supercell.

There are two main differences between diamond and SiC. Firstly, SiC has two inequivalent substitutional sites, i.e. the silicon site and carbon site. Despite of the same symmetry, the wave function of nickel's $3d$ electron will have a smaller overlap with NN carbon's dangling bond than NN silicon's dangling bond. Compared with diamond, the larger size of a silicon atom suppresses the spreading of nickel's electronic wave function while the looser relaxation space of nickel's central tetrahedron provides an additional tensile hydrostatic strain. Therefore, we expect this spin center to exhibit a higher transition strain than Ni$_\textrm{s}$ in diamond\cite{Chanier2012}, especially C-sub Ni in SiC.


Secondly, Si and C both have four electrons in the outmost subshell and share the same symmetry in the SiC matrix. Nevertheless, the difference in Si and C's electronegativities leads to a charge transfer between them. Before calculating the impurity-contained SiC supercell, the pure 4H and 3C SiC supercells are checked with FP-LOMB, which is implemented in FPLO package\cite{Koepernik1999}, to investigate the charge transfer between Si and C. Note that WIEN2K is implemented with LAPW method so that it's unable to evaluate the precise occupation information. The calculated Si$\rightarrow$C charge transfers are shown in the following table:
\begin{table}[h]
\centering
\begin{tabular}{ccccc}
\hline
\hline
3C-SiC & &  & & \\ 
Atom & & Q-tot & & Q-excess\\ \hline
C & & 7.032 & & 1.032 \\ \hline
Si  & & 12.968 & & -1.032  \\ \hline
\hline
4H-SiC & &  & & \\ 
Atom & & Q-tot & & Q-excess\\ \hline
C & & 7.053 & & 1.053 \\ \hline
Si  & & 12.947 & & -1.053  \\ \hline
\hline
\end{tabular}
\caption{Si$\rightarrow$C charge transfers in 3C and 4H SiC.}
\end{table}
It can be seen that the silicon atoms give away 1.03-1.05 electrons/atom to carbon atoms so that the ionic formula for SiC becomes Si$^\textrm{+1.04}$C$^\textrm{-1.04}$. This result is 0.3 electrons/atom off that in \cite{Zhao2000}, which is due to the difference in functionals we use. The charge transfer mainly occurs in the formation of Si-C bond, which is about 0.26 electrons/bond. The excess charge in carbon atoms makes the difference between local environment of impurities in diamond and SiC not only the atomic separation, but the on-site coulomb potential.

\section{Results}
\subsection{Carbon substituted nickel in 3C silicon carbide}
\label{SiC_c}
With above setup, the C-sub Ni$_\textrm{s}$ is investigated. Considering the odd number occupation of $3d$ electron in nickel, we investigated the effect of Jahn-Teller distortion first by allowing a P1 symmetric atomic relaxation. Compared with $T_d$ symmetric relaxation, only a difference of 0.09 eV in the total energy is observed. For simplicity, the relaxation is constrained to $T_d$ symmetry in C-sub case. Under ambient pressure, the whole unit cell is anti-ferromagnetic (AFM). As the lattice constant is extended by 6\%, the antiferromagnetic-ferromagnetic transition shows up, which is similar to the result in \cite{Chanier2012}.

This bonding mechanism between Ni and nearest neighbor (NN) Si can be explained via the $p-d$ hybridization model. Under the crystal field, the degeneracy of $3d$ is lowered and splitted into $t_2$ and $e$ orbitals, which have a degeneracy of three and two respectively. $t_2$ orbital approaches the silicon along the Si-Ni axis which can form sigma bonds with ligands so that it locates above the e orbitals. The dangling bonds with $3s3p$ character are splitted into $a_1$ and $t_2$ level, which can hybridize with the $3d$ derived $t_2$ level. The orbital partial occupation is listed as follows:

\begin{table}[h]
\centering
\begin{tabular}{cccccccccccccccc}
\hline
\hline
 & & $3s$ & & $3p$ & & $4s$ & & $5s$ & & $3d$ & & $4d$ & & $4p$ \\ \hline
Ni$_\textrm{s}$ & & 2.000 & & 6.007 & & 0.628 & & -0.003 & & 8.731 & & 0.106 & & 0.518 \\ \hline
Ni$_\textrm{s}$ spin & & 0 & & 0 & & 0.015 & & 0.001 & & 0.306 & & -0.008 & & 0.202 \\ \hline
\hline
& & $2s$ & & $2p$ & & $3s$ & & $4s$ & & $3p$ & & $4p$ & & $3d$ \\ \hline
NN $Si$ & & 2.000 & & 6.000 & & 1.149 & & -0.009 & & 2.056 & & -0.022 & & 0.369 \\ \hline
NN $Si$ spin & & 0 & & 0 & & 0.033 & & 0 & & 0.178 & & 0.002 & & 0.026 \\ \hline
\hline
\end{tabular}
\caption{Projected occupation and spin of Ni$_\textrm{s}$ and NN silicon.}
\end{table}

Before looking into the detailed population and spin configuration, we need to note that the occupation here is a result of hybridization between levels with the same symmetry. In other words, it cannot offer a precise description on the electron configuration before hybridization. To recover the pre-hybridized configuration, we need to sort the orbitals according to their symmetries.

The closed shell electrons, including $3s_\textrm{Ni}$, $3p_\textrm{Ni}$, $2s_\textrm{Si}$ and $2p_\textrm{Si}$, won't participate the formation of bonds. The $4s_\textrm{Ni}$ and $3s_\textrm{Si}$ give rise to $a_1$ hybrid orbital, while the $3d_\textrm{Ni}$, $3p_\textrm{Si}$, $4p_\textrm{Ni}$ and $3d_\textrm{Si}$ form $t_2$ level. Note that part of the $3d_\textrm{Ni}$ electrons will be splitted into an $e$ level due to the crystal field as before. As a result, 1.76($\approx$2) electrons and 7.677($\approx$8) electrons will occupy the 4s/3s derived $a_1$ and $3d_\textrm{Ni}$/$4p_\textrm{Ni}$/$3p_\textrm{Si}$/$3d_\textrm{Si}$ derived $t_2$ level respectively. Regarding the $a_1$ level, both of the $4s_\textrm{Ni}$ and $3s_\textrm{Si}$ electrons are spinless, which indicates the full occupation of the bonding and anti-bonding $a_1$. As for the $t_2$ level, we can take the Ni atom as a reference. Ni typically exhibits a configuration of $3d^9 4s^1$. Both of the $4p_\textrm{Ni}$ and $3d_\textrm{Si}$ levels are far above the outermost electron in Ni and Si atoms. Although they are partially occupied, they are still supposed to be empty before hybridization. Therefore, the electron assignments will be $3d^9$ and $3p^3$, in which $3p^3$ exhibits $\uparrow\uparrow\downarrow$ configuration to achieve a polarized subshell. Here, the unquantized $d$ shell occupation can actually be attributed to the mixing of $3d^9$ and $3d^8$ as the result of transition metals on the Cu$_2$N/Cu(100) interface \cite{PhysRevB.92.174407}.

As will be discussed later, Si-sub and C-sub Ni$_\textrm{s}$ essentially have the same bonding mechanism while the C-sub case has a higher occupation in Ni site. The main reason for this difference is the polarization of $4s$ orbital between Ni$_\textrm{s}$ and NN silicon. In spite of the same configuration, $4s$ orbital is less extensive due to the lower electronegativity of silicon so that Ni$_\textrm{s}$ will appear as a neutral atom. Therefore, both of dangling bond and Ni$_\textrm{s}$ have a net magnetic moment of 1 $\mu_B$. Other than that, the deviation of silicon's occupation from neutral indicates the polarization in Si-C bonds.

\begin{figure}
    \begin{subfigure}[b]{0.5\textwidth}
        \includegraphics[width=\textwidth]{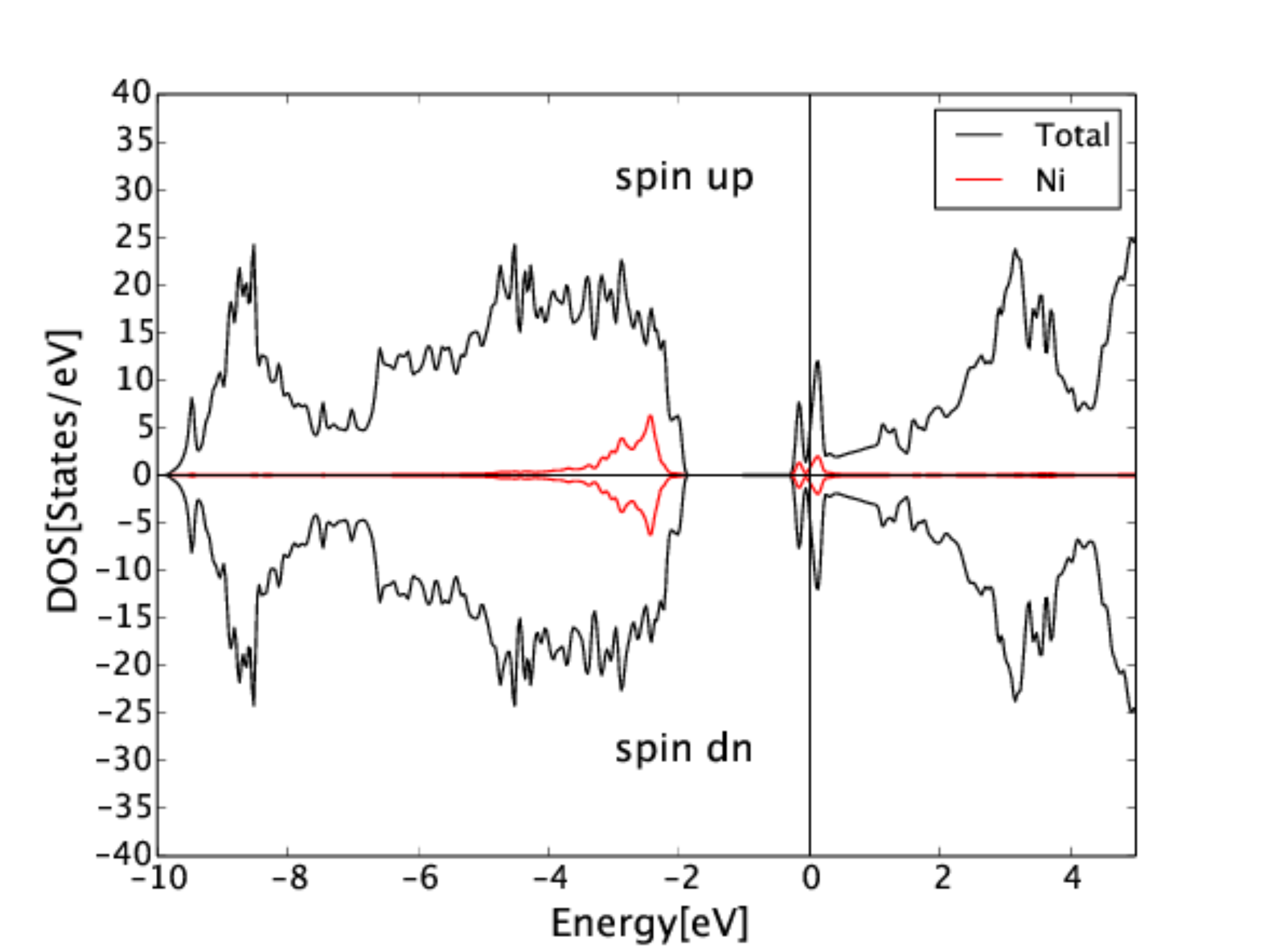}
        \caption{}
        \label{fig:4strainDOSCsub}
    \end{subfigure}
    ~ 
    \begin{subfigure}[b]{0.5\textwidth}
        \includegraphics[width=\textwidth]{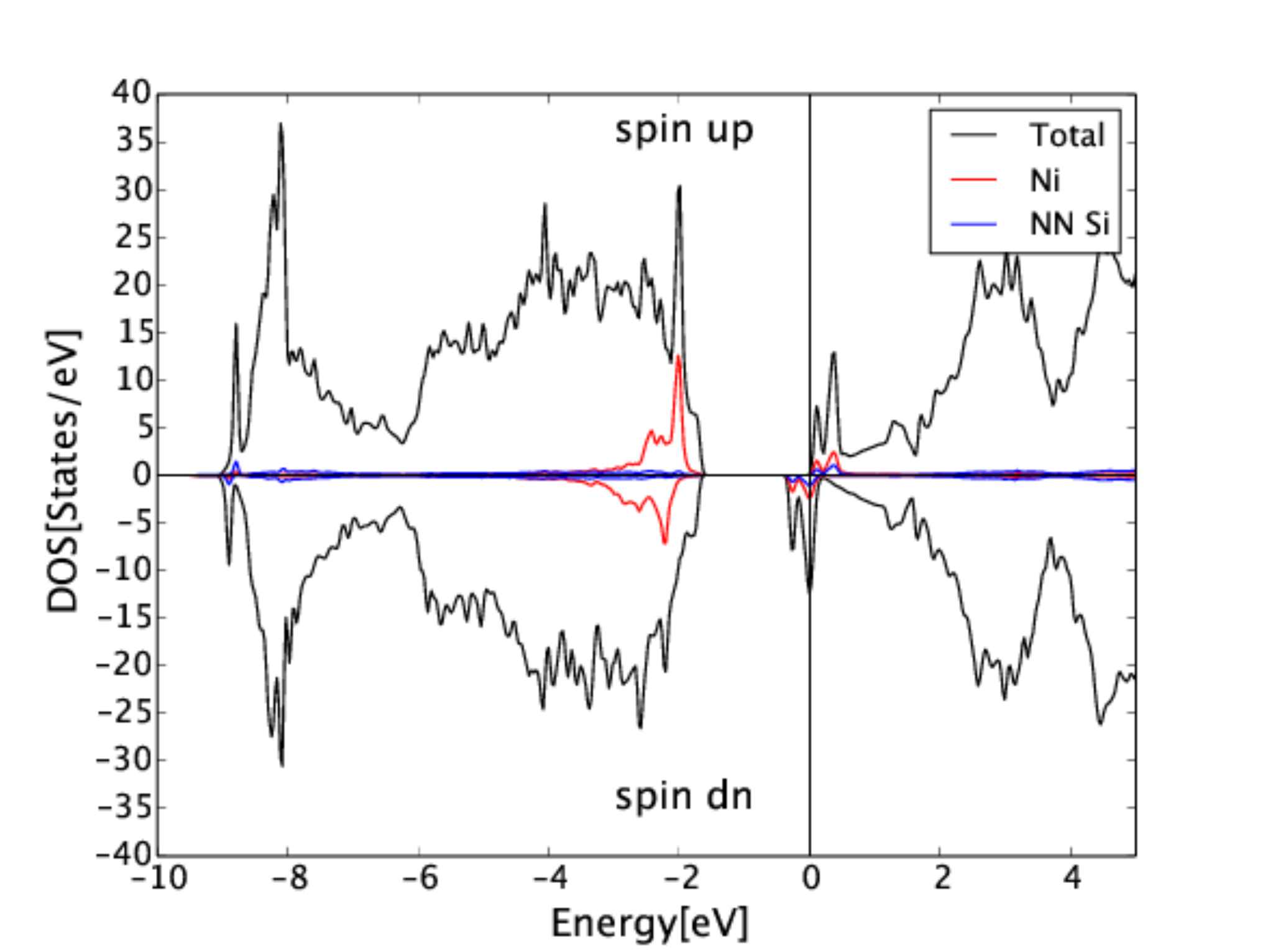}
        \caption{}
        \label{fig:8strainDOSCsub}
    \end{subfigure}
    \caption{Total density of states and partial density of states of nickel and nearest-neighbor silicon of C-sub case under 4\% (a) and 8\% (b) tensile strain.}
\end{figure}

As shown in Fig.\ref{fig:4strainDOSCsub}, we can see the total and nickel's partial DOS under 4\% strain. Due to the absence of a magnetic moment, a symmetric spin configuration can be observed straightforwardly. Near the conduction band edge, we can see a partially occupied antibonding level locating at 2 eV above the valence band maximum (VBM), which is attributed to Ni's $3d$ electron and dangling bonds.  About 0.5 eV below the VBM, $e$ levels can be observed. Here, the defect level adjoins and extends into conduction band, which is due to the underestimation of the SGGA functional. The calculated bandgap of defect-free 3C-SiC under ambient strain is 1.362 eV compared with the experimental value of 2.36 eV.

For comparison, we show the total and nickel's partial DOS under 8\% strain in Fig.\ref{fig:8strainDOSCsub}. The electronic structure exhibits Stoner behaviour with an asymmetry in the spin-up and spin-down channels. Around 2 eV and 1.8 eV above VBM, two anti-bonding levels are formed by the $2p^2_\textrm{Si}$ and $3d^{10}_\textrm{Ni}$ derived $t_2$ levels, of which one is partially occupied and the other one is empty. The two spin-up anti-bonding electrons render the whole unit cell ferromagnetic. Again, we can see two $3d$ derived $e$ levels around 0.3 eV below the VBM, while they are separated by a Hund exchange splitting $J=0.1$ eV. 

\begin{figure}[h!]
    \includegraphics[width=0.5\textwidth]{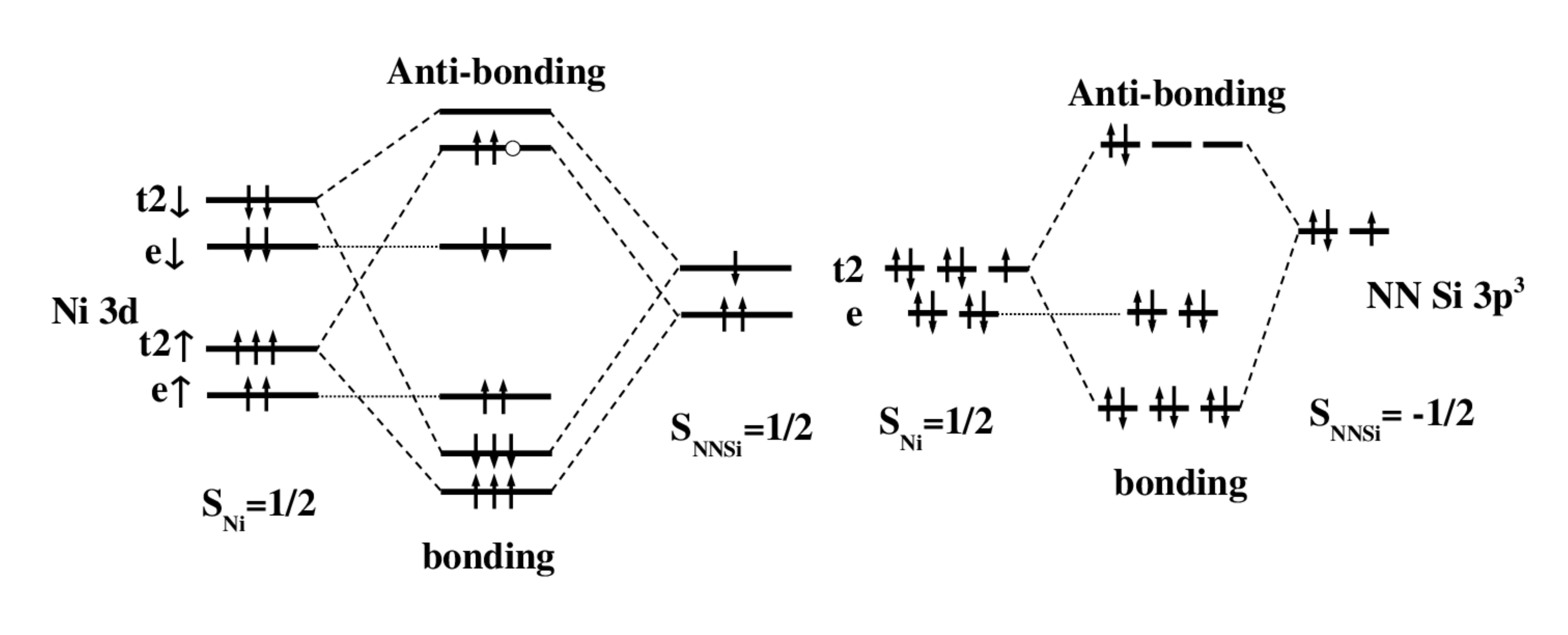}
  \caption{Hybridization between Ni and NN silicon in FM(left) and AFM(right) phase.}
\label{CsubHyb}
\end{figure}

\begin{figure}[h!]
  \centering
    \includegraphics[width=0.5\textwidth]{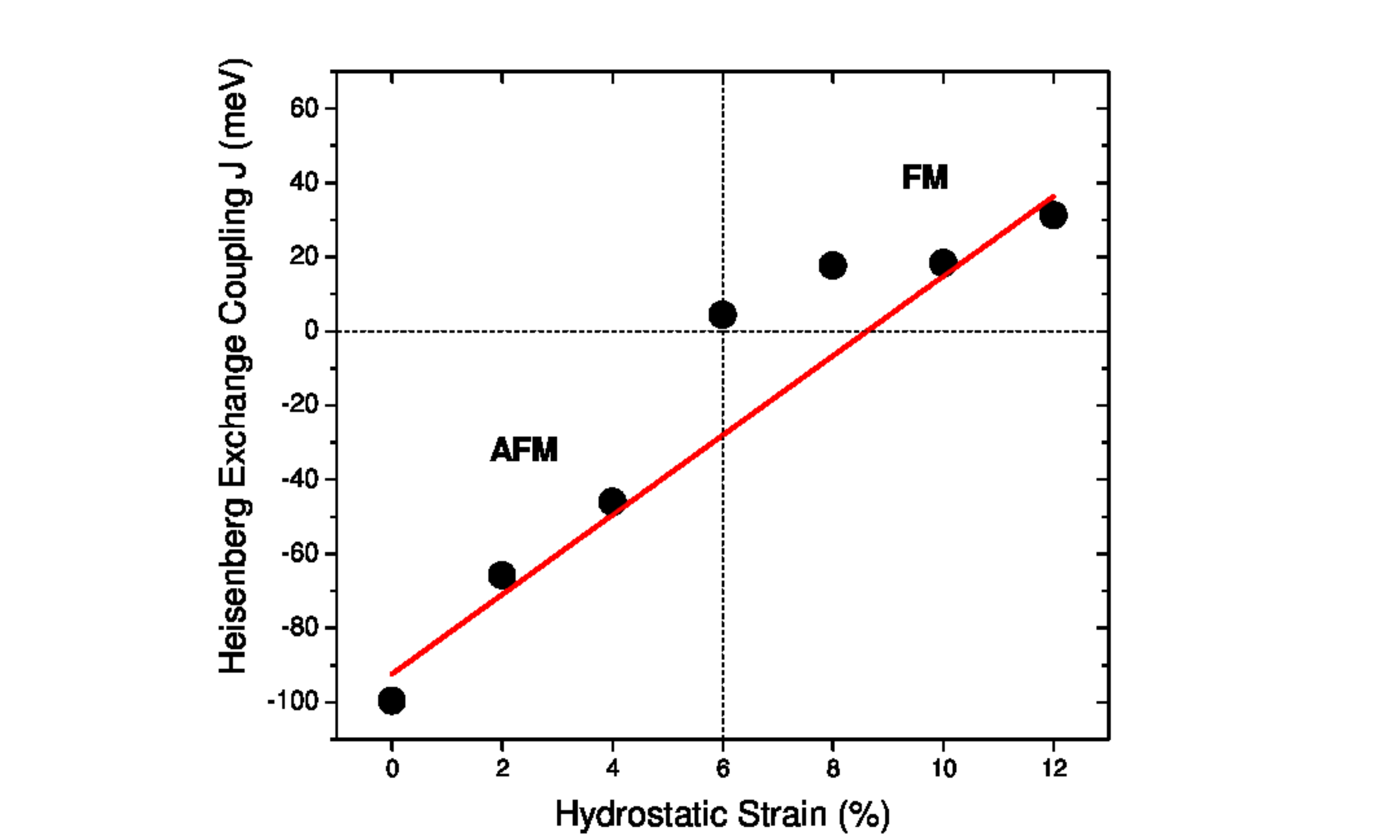}
  \caption{Heisenberg exchange between the two spins localized to nickel and surrounding silicons in the C-sub case.}
\label{Heisenberg3cCsub}
\end{figure}

The bonding mechanism for the AFM and FM phases are summarized with a diagram in Fig.\ref{CsubHyb}. For simplicity, we neglect some levels that participate in the hybridization but only make minor contributions. Apparently, the spin splitting in $3d$ level derived crystal field levels will only happen in the FM phase but not the AFM phase. This diagram also identifies the origin of the ferromagnetism, namely the exchange interaction between the $\frac{1}{2}$-spin residing in dangling bond and that in nickel transfer from dangling bond. The Heisenberg exchange coupling $J_H=(E_{AFM}-E_{FM})$ between these neighboring $\frac{1}{2}$ spins stabilizes the parallel alignment. The strain-dependent exchange coupling energy is evaluated and shown in Fig.\ref{Heisenberg3cCsub}. To calculate the total energy of metastable states with a magnetic phase different from ground states, we made constrained DFT calculations with the magnetic moment fixed to the value of interest. The total energy is minimized with the constrained magnetic moment. The exchange coupling energy varies linearly as the strain with a slope of 11.2 meV/\%. Note that the Heisenberg exchange energy per atom is within 10 meV. In the diluted magnetic semiconductor, the atomic disorder of the host atom and impurity caused by unnegligible entropy will cause the reversal of magnetic states at finite temperature\cite{Los2007}. However, the result we achieve here is from a single dopant rather than diluted doped, which means the group disorder won't have an impact on the single impurity. 

To describe the C-sub  Ni$_\textrm{s}$ spin center, we treat it as a system consists of two local $\frac{1}{2}$-spins, in which the coupling strength can be manipulated with strain. The spin Hamiltonian can be written as
\begin{equation}
H=-2J \textbf{S}_\textrm{Ni} \cdot \textbf{S}_\textrm{d}+ g_\textrm{Ni} \mu_\textrm{Ni} B_z  S_\textrm{Niz}+ g_\textrm{d} \mu_\textrm{d} B_z  S_\textrm{dz}.
\end{equation}
Here, the external magnetic field $B_z$ splits off the $T_{+/-}$ so as to orient the magnetic moment of unit cell. By selecting the triplet $T_0$($s=1$, $s_z=0$) and singlet $S$ as the logical qubits, we can create a decoherence-free space immune from collective dephasing. In the representation of $T_0$/$S$, the spin Hamiltonian can be written as
\begin{equation}
\label{manipulation}
H=\begin{bmatrix}
	-J/2 & \mu_B B_z \Delta g/2 \\
	\mu_B B_z \Delta g/2 & J/2
\end{bmatrix},
\end{equation}
where $\Delta g=g_\textrm{Ni}-g_\textrm{d}$. 

Due to the delocalizaton of the dangling bond, we can use the g-factor of free electron to describe it. As for nickel, we need to know the total angular momentum and orbital angular momentum of the holes in d shell. In reality, there is a 	predominant spin-orbital interaction in nickel which splits the 3d derived $t_2$ further. $t_2$ can be spanned in the space of $d_{xz}$, $d_{yz}$ and $d_{xy}$, which can form orbitals with $m=+1,0,-1$. Note that the $m=0$ orbital is a mix of $m=2$ and $m=-2$ orbitals. Due to the spin-orbital coupling, the holes occupy the highest level which is parallel to the spin. We can get the total angular momentum $J=3/2$ and $g_1=1.33$. 

\begin{figure}[h!]
  \centering
    \includegraphics[width=0.5\textwidth]{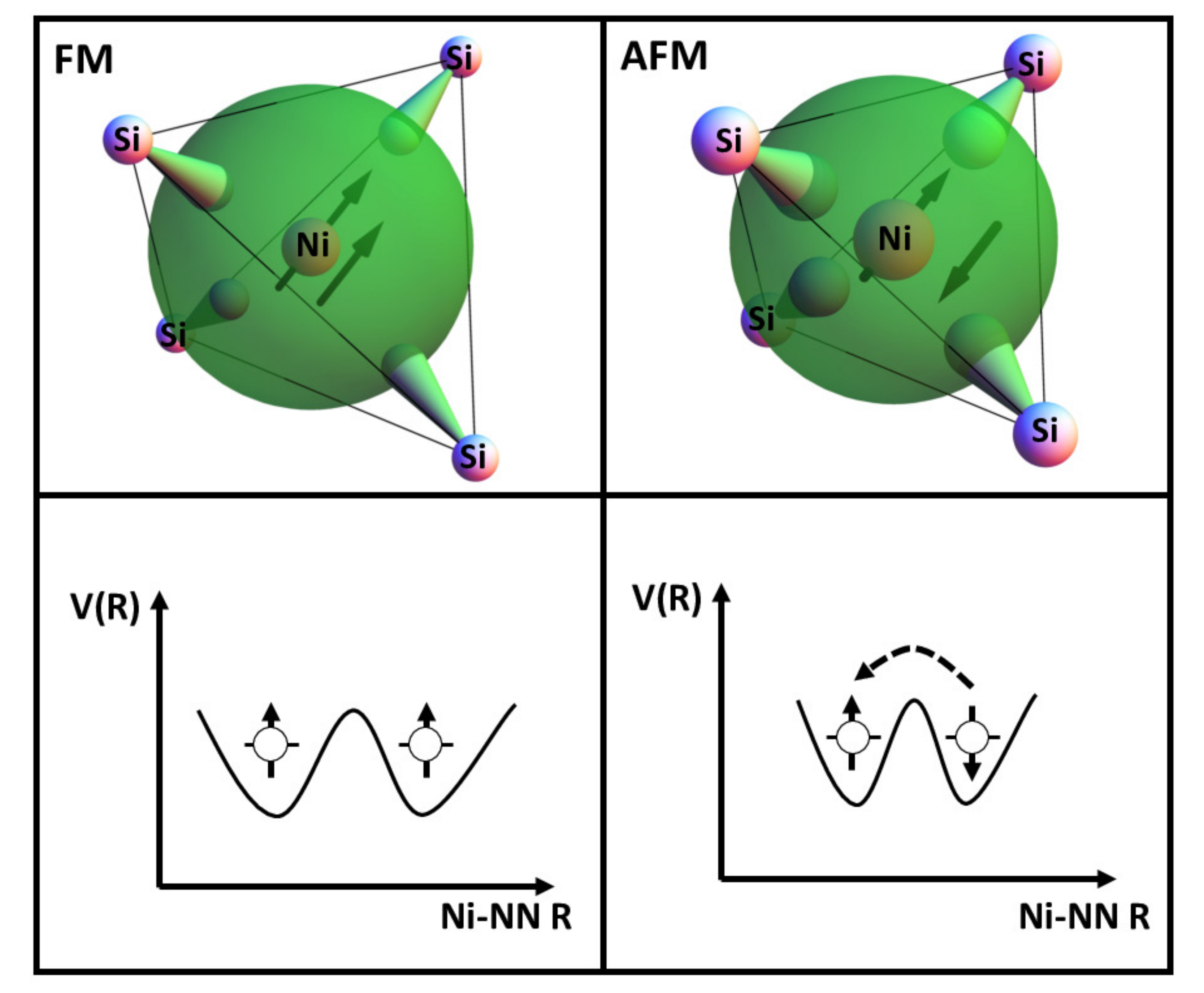}
  \caption{Virtual hopping induced AFM-FM transition due to the Ni-NN distance.}
  \label{potentialwell}
\end{figure}

In Chanier $et\ al.$'s work, they proposed a model of Hund's rule driven hopping between the pair of spins in Ni and the dangling bond, which is also applicable in our problem\cite{Chanier2012}. As shown in Fig.\ref{potentialwell}, the varying Ni-NN distance due to the applied strain change the probability of virtual hopping between nickel and NN silicon. According to double-exchange model, a spin possesses a lower kinetic energy if it can hop between two potential wells without changing its spin direction. This contribution to the exchange energy depends on the hopping and on-site coulomb interaction:
\begin{equation}
E_{hop}\propto -\frac{t^2}{U}.
\end{equation} 
The hopping probability inversely depends on the spatial separation $R$ and energy separation $W$ between the occupied states\cite{Mott1968}
\begin{equation}
P\sim exp[-2\alpha R-\frac{W}{kT}].
\end{equation}
Therefore, the varying Ni-NN distance effects the $3d_\textrm{Ni}$-$sp_\textrm{NN}$ hopping strength so as to induce this AFM-FM transition. 

\subsection{Silicon substituted nickel in 3C silicon carbide}
\label{SiC_si}
The strain effect on the magnetic states of Si-sub nickel in 3C-SiC is also investigated. The $P1$ symmetry relaxation check shows that the total energy is only 0.5 eV stabler than that under $T_d$ symmetry due to Jahn-Teller distortion. NN carbon is distorted closer to the impurity by 0.003 $\AA$. As a result, the structure optimization will obey $T_d$ symmetry only. Hydrostatic strain from compressive 15\% to tensile 2\% have been considered. Due to the similar size of nickel and silicon, the NN carbons are pushed out along diagonal direction by only 1\%. Compared with Ni$_\textrm{s}$ in diamond(14\%), this results implies a much lower formation energy. Also, this equivalent built-in tensile strain indicates a much larger strain is required to realize the antiferromagnetic(AFM)-ferromagnetic (FM) transition. 

\begin{figure}[h!]
  \centering
    \includegraphics[width=0.5\textwidth]{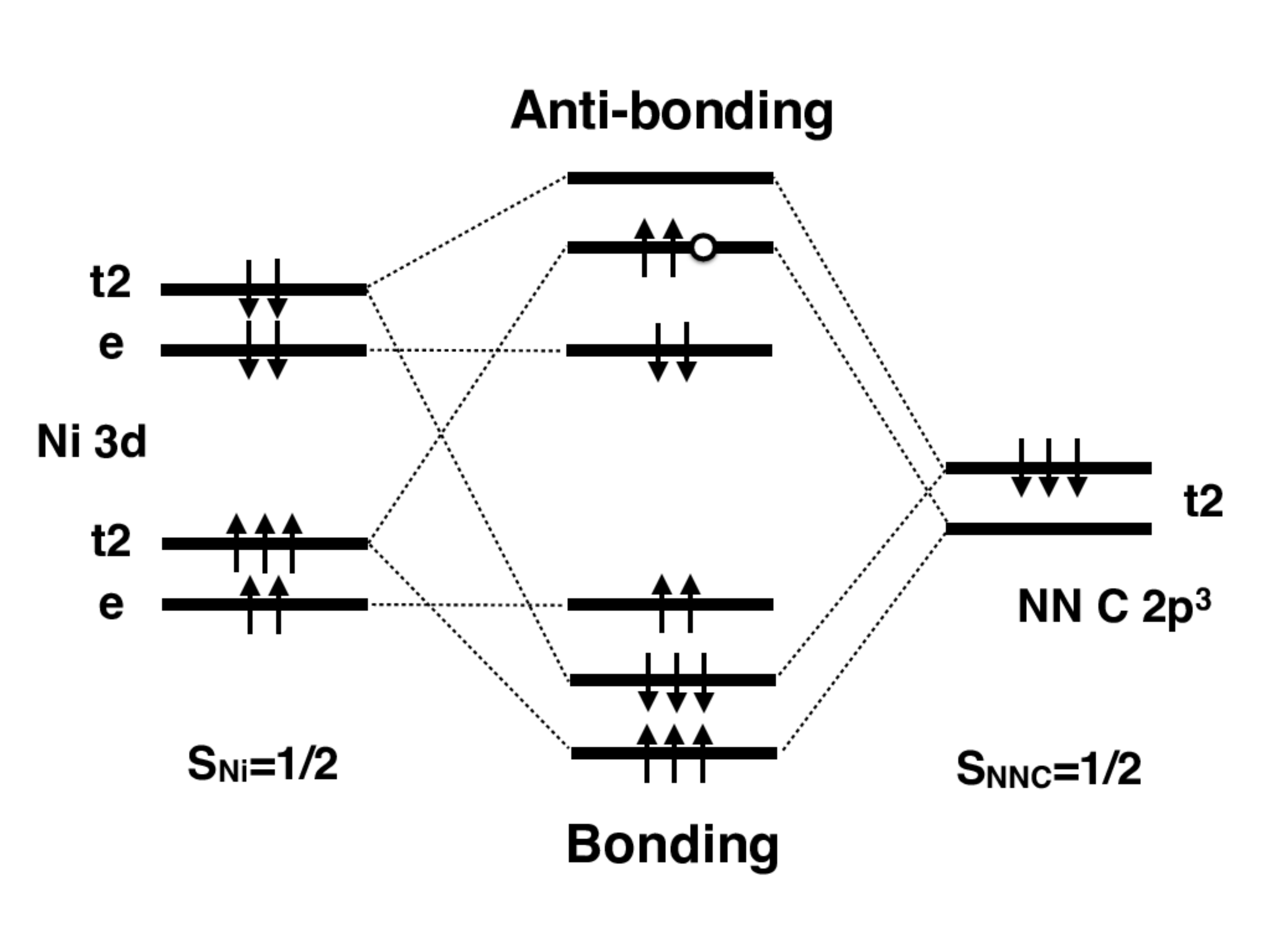}
  \caption{Hybridization between Ni's 3d electrons and NN carbon's dangling bond.}
\label{SisubHyb}
\end{figure}

Similarly, the $p$-$d$ hybridization model can be used to describe our calculation. We calculated the total and projected occupation of nickel and NN carbon with the method of FP-LOMB. The occupations are shown in the Table.\ref{C-sub-occ}. As a result of the big difference in the electronegativity between C and Ni, the outmost $4s$ electron of Ni is attracted by the dangling bond of NN carbon, which is verified by the total occupation of 27.2 for Ni$_\textrm{s}$. Consequently, Ni$_\textrm{s}$ exhibits an electron configuration of $3d^9 4s^0$ while the dangling bond is in $2s^2 2p^3$ configuration. Before the dangling bond hybridize with Ni$_\textrm{s}$, there is a 1 $\mu_B$ net magnetic moment residing in both of them. After the crystal field splitting, the derived level is as shown in Fig.\ref{SisubHyb}. With the same symmetry, the $3d_\textrm{Ni}$ derived $t_2$ hybridizes with $2p_{NN}$ derived $t_2$ to form an anti-bonding level ($t_{AB}$) and bonding ($t_B$) level. 
 
\begin{table}[h]
\centering
\begin{tabular}{cccccccccccccccc}
\hline
\hline
 & & $3s$ & & $3p$ & & $4s$ & & $5s$ & & $3d$ & & $4d$ & & $4p$ \\ \hline
Ni$_\textrm{s}$ & & 1.999 & & 6.004 & & 0.452 & & -0.010 & & 8.421 & & 0.096 & & 0.522 \\ \hline
Ni$_\textrm{s}$ spin & & 0 & & -0.003 & & 0.012 & & 0 & & 0.879 & & -0.016 & & 0.116 \\ \hline
& & $1s$ & & $2s$ & & $3s$ & & $2p$ & & $3p$ & & $3d$ & & \\ \hline
NN $C$ & & 2.000 & & 1.374 & & 0.001 & & 3.215 & & -0.003 & & 0.023 \\ \hline
NN $C$ spin & & 0 & & 0.014 & & 0 & & 0.159 & & 0.001 & & 0.002 \\ \hline
\hline
\end{tabular}
\caption{Projected occupations of Ni$_\textrm{s}$ and NN carbon.}
\label{C-sub-occ}
\end{table}

\begin{figure}[h!]
  \centering
    \includegraphics[width=0.5\textwidth]{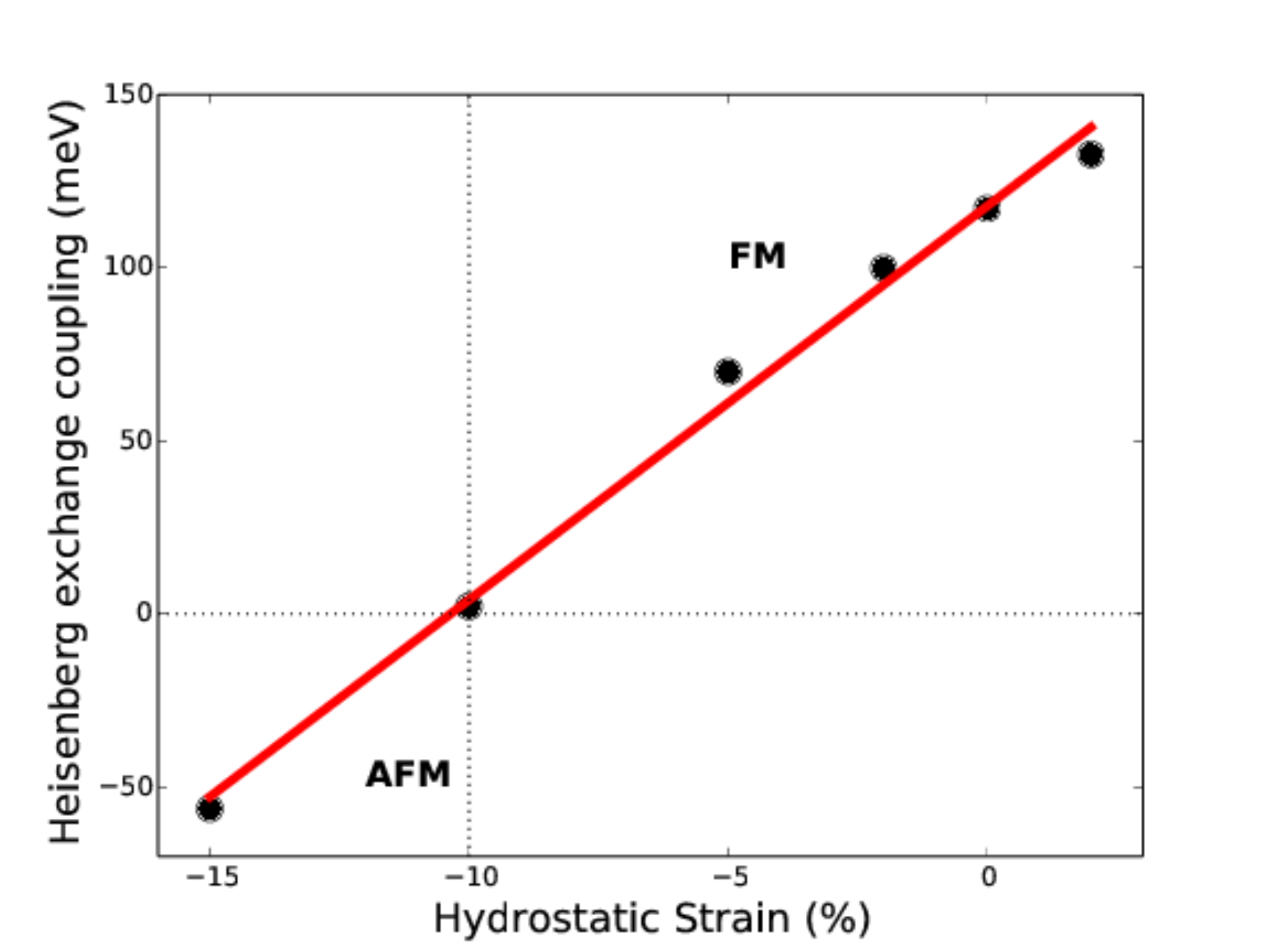}
  \caption{Heisenberg exchange between the two spins localized to Ni and surrounding carbons in the Si-sub case.}
\label{Heisenberg3c}
\end{figure}

The dependence of $J$ on the hydrostatic strain is shown in Fig.\ref{Heisenberg3c}. As the compressive strain is raised above 10\%, an AFM-FM transition can be observed. According to the stiffness tensor of 3C-SiC\cite{stiffness}, this strain is equivalent to a pressure $\sim$ 114 GPa, which can be achieved by using diamond anvil cell\cite{Dubrovinsky2015}. This result also confirms our prediction of a higher transition strain since the Ni-NN spatial separation is larger in SiC. To initialize qubits, a compressive strain can be used to prepare it in the $T_+$ since it's in the FM phase. Then a microwave radiation can coherently manipulate the spin into $T_0$ state. After initialization, we can realize arbitrary one-qubit operations by strain-based and applied magnetic field according to Eq. \ref{manipulation}. 

\subsection{Nickel in 4H silicon carbide}
\label{4HSiC}
After discussing on cubic SiC, we want to look into one of the most important hexagonal SiC polytypes, namely 4H-SiC. There are two inequivalent Si and C sites in 4H-SiC as shown in Fig.\ref{SiCstack}. However, the substitutional sites has minor impact on the electronic property\cite{Miao2006} due to the similarity in their local environment so that only h site is considered in our work. The applied hydrostatic strain is adjusted from -10\% to +6\%. To achieve an accuracy of 10 meV, a $4\times 4\times 4$ Monkhorst-Pack k-mesh is used.


The AFM and FM solutions are obtained through magnetic moment constrained calculation. The discrepancy between the AFM and FM phases (within 100 meV) shows that the Heisenberg exchange coupling is not sensitive to the strain and close to the thermal energy at room temperature. In thermodynamic equilibrium, its statistical distribution can be described by Maxwell-Boltzmann equation:
\begin{equation}
\frac{n_{FM}}{n_{AFM}}=e^{-\frac{E_{H}}{kT}},
\end{equation}
where the $E_{H}$ is the Heisenberg exchange energy. This ratio is around 0.607 at room temperature, which implies that AFM and FM states can exist concurrently. Except for the AFM ground state, we also observed an unstable magnetic local minimum (total magnetic moment of 0.6 $\mu_B$) with tensil strain, which should be a mixed state of AFM and FM phases. A nonzero tunnelling probability between these configurations gives rise to this noninteger magnetic moment, which is similar to the anomalous thermal properties of glasses below 1K\cite{Anderson1972}. The lower-energy curve are both the nonmagntic solution while the higher-energy are both the magnetic solution. 

\begin{figure}[h!]
  \centering
    \includegraphics[width=0.5\textwidth]{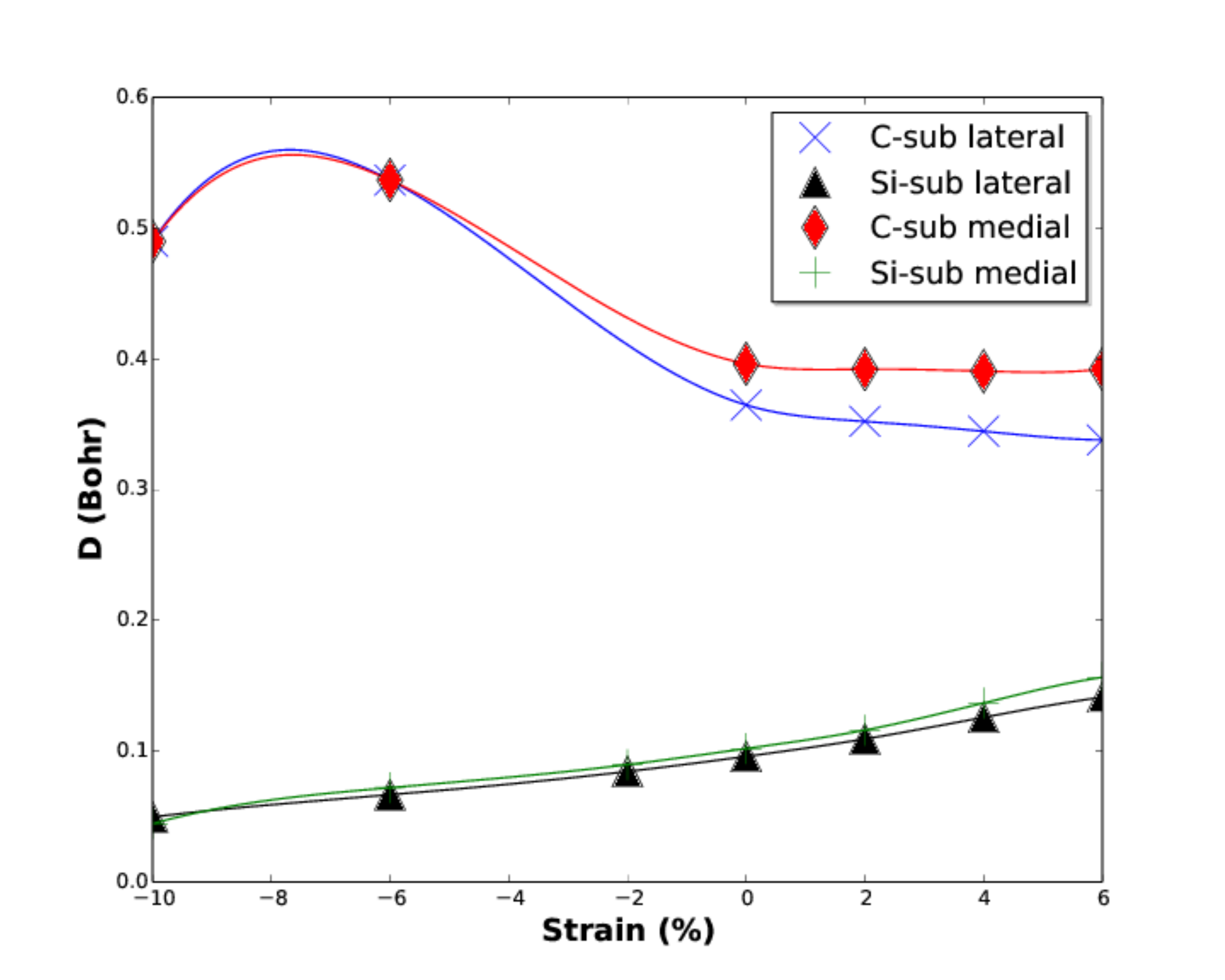}
  \caption{Volume dependence of Ni-NN relaxation in ferromagnetic and anti-ferromagnetic states.}
  \label{4HNNrelaxation}
\end{figure}

After atomic position optimization, we can see that the C-sub case is relaxed to a much larger extent compared with Si-sub case (Fig.\ref{4HNNrelaxation}). Due to that the radius of Ni and Si are similar, the Ni-NN distance increases linearly in Si-sub case. In contrast, the relaxation of Ni-NN distance from the undoped and strained SiC:
\begin{equation}
D=d_\textrm{Ni-NN}-d_\textrm{Si-C}
\end{equation}
reaches the maximum at -8\% to -6\% compressive strain. Compared with the lateral NN atom, the medial NN atom in both cases have a larger relaxation, whose unpaired $p_z$ electron gives rise to the CFS $a_1$ as shown in Fig.\ref{SiCstack}. This pattern agrees well with the level configuration derived  from the group theory. 

The detailed population of Ni and its NN atoms via LOMB method are listed in the table \ref{4Hocc}. Due to the charge transfer between silicon and carbon atom, nickel carries a more accurate information here. The hybridized $3d$ shell of nickel consists of $3d$, $4d$ and $4p$ orbitals, which gives a total $3d$ occupation of 9.089 and 9.454 in the Si-sub and C-sub cases. These orbitals form $a_1$ and $e$ hybridized levels ($d_{z^2}\rightarrow a_1$, \{\{$d_{xz}$,$d_{yz}$\},\{$d_{xy}$,$d_{x^2-y^2}$\}\}$\rightarrow e$) together with NN-Si/C's $3s3p/2s2p$, which have the same symmetry. Other than that, the occupation of $3d$ shell indicates that the nickel is in the +1 valence state in Si-sub case while the mixed state of +1 and neutral state in C-sub case. 

\begin{table}[h]
\centering
\begin{tabular}{cccccccccccccccc}
\hline
\hline
Si-sub & &  & &  & &  & &  & &  & &  & &  \\ \hline
Ni$_\textrm{s}$ & & $3s$ & & $3p$ & & $4s$ & & $5s$ & & $3d$ & & $4d$ & & $4p$ \\ \hline
occ. & & 1.999 & &   6.006  & &  0.396  & &  -0.010  & &  8.463  & &  0.103  & &  0.523 \\ \hline
spin & & 0.000  & &  0.000 & &  0.000 & &  0.000 & &  0.000  & &  0.000 & &  0.000 \\ \hline
C$_\textrm{3}$ & & $1s$  & &  $2s$  & &  $3s$  & &  $2p$  & &  $3p$  & &  $3d$ \\ \hline
occ. & & 2.000 & & 1.355 & & 0.003 & & 3.248 & & 0.013 & & 0.027 \\ \hline
spin & & 0.000 & & 0.000 & & 0.000 & & 0.000 & & 0.000 & & 0.000 \\ \hline
C$_\textrm{1}$ & & $1s$  & &  $2s$  & &  $3s$  & &  $2p$  & &  $3p$  & &  $3d$ \\ \hline
occ. & & 2.000 & & 1.331 & & 0.002 & & 3.266 & & 0.005 & & 0.027 \\ \hline
spin spin & & 0.000 & & 0.000 & & 0.000 & & 0.000 & & 0.000 & & 0.000 \\ \hline
\hline
C-sub & &  & &  & &  & &  & &  & &  & &  \\ \hline
Ni$_\textrm{s}$ & & $3s$ & & $3p$ & & $4s$ & & $5s$ & & $3d$ & & $4d$ & & $4p$ \\ \hline
occ. & & 1.999 & & 6.013 & & 0.560 & & -0.002 & & 8.781 & & 0.128 & & 0.545 \\ \hline
spin & & 0.000 & & 0.000 & & 0.000 & & 0.000 & & 0.000 & & 0.000 \\ \hline
Si$_\textrm{3}$ & & $2s$ & & $2p$ & & $3s$ & & $4s$ & & $3p$ & & $4p$ & & $3d$ \\ \hline
occ. & & 1.999 & & 6.000 & & 0.978 & & -0.009 & & 2.067 & & -0.019 & & 0.513 \\ \hline
spin & & 0.000 & & 0.000 & & 0.000 & & 0.000 & & 0.000 & & 0.000 & & 0.000  \\ \hline
Si$_\textrm{1}$ & & $2s$ & & $2p$ & & $3s$ & & $4s$ & & $3p$ & & $4p$ & & $3d$ \\ \hline
occ. & & 1.999 & & 6.000 & & 0.975 & & -0.010 & & 2.033 & & -0.020 & & 0.495  \\ \hline
spin & & 0.000 & & 0.000 & & 0.000 & & 0.000 & & 0.000 & & 0.000 & & 0.000  \\ \hline
\hline
\end{tabular}
\caption{Projected occupations of Ni$_\textrm{s}$ and NN atoms in Si-sub and C-sub cases.}
\label{4Hocc}
\end{table}

\begin{figure}
    \begin{subfigure}[b]{0.5\textwidth}
        \includegraphics[width=\textwidth]{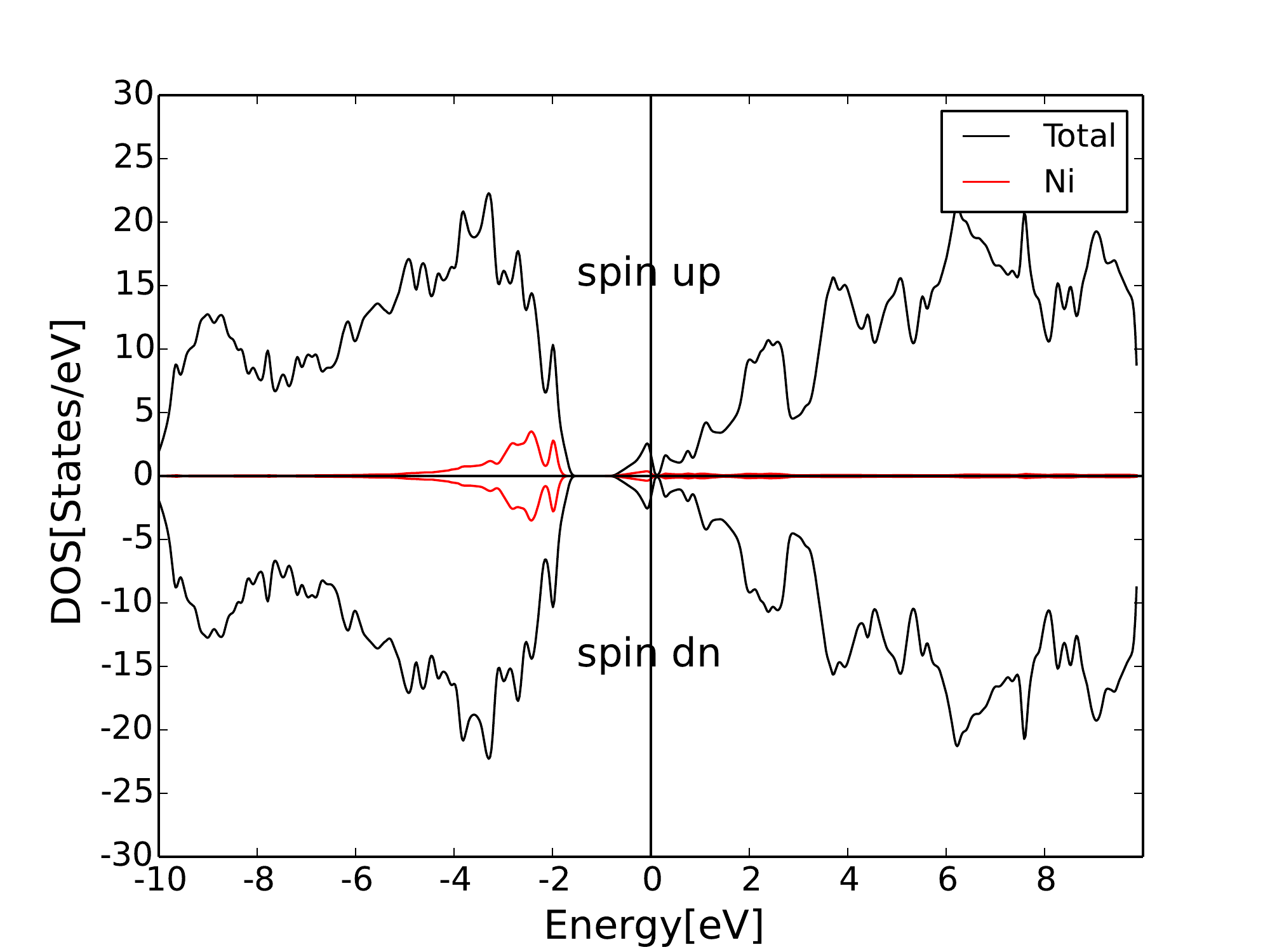}
        \caption{}
        \label{fig:4HCDOS}
    \end{subfigure}
    ~ 
    \begin{subfigure}[b]{0.5\textwidth}
        \includegraphics[width=\textwidth]{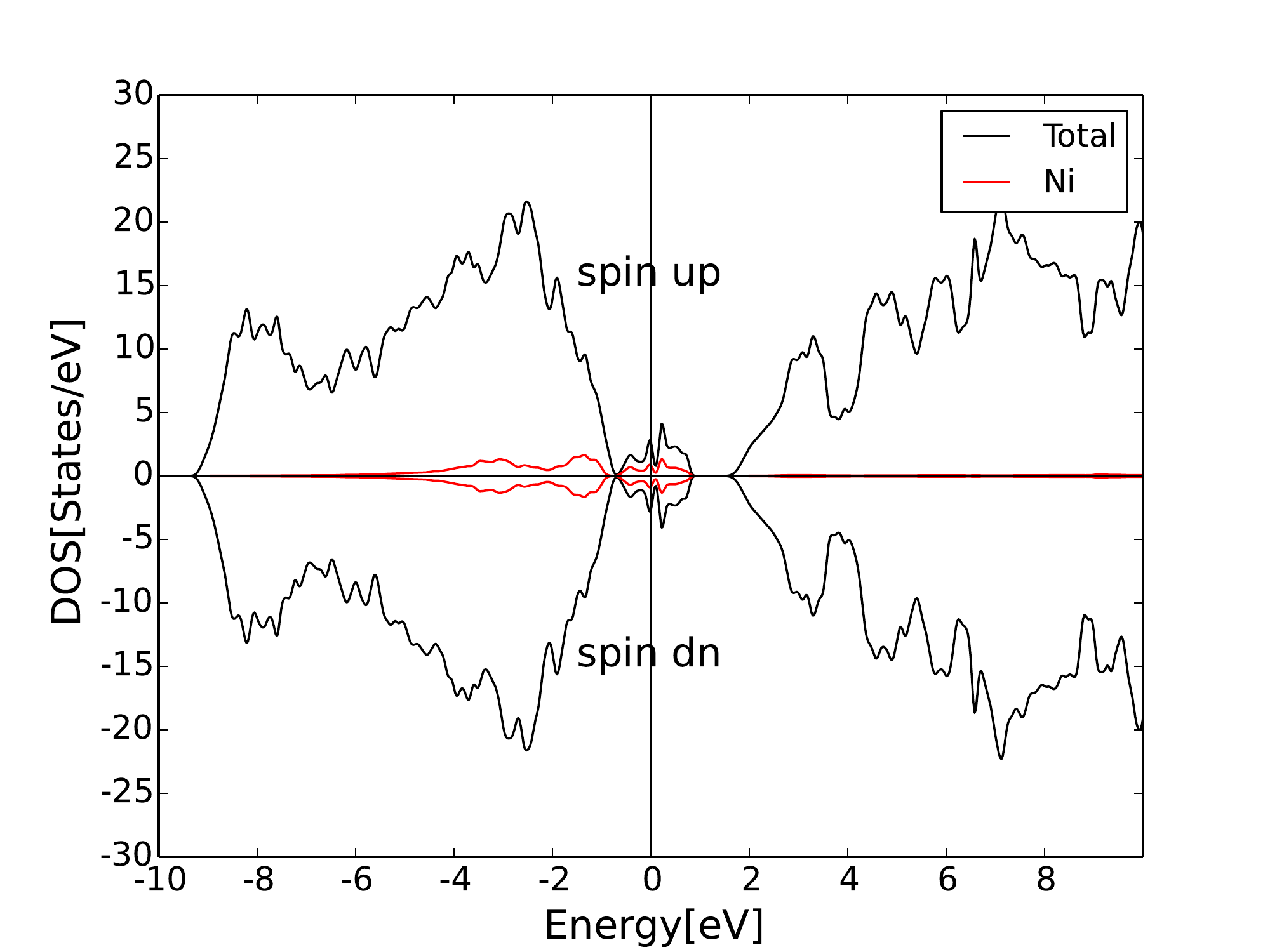}
        \caption{}
        \label{fig:4HSiDOS}
    \end{subfigure}
    \caption{DOS of C-sub (a) and Si-sub (b) Ni in 4H-SiC.}
    \label{4HDOS}
\end{figure}

To verify the electronic configuration of impurity and complex, the total and partial density of states in C and Si substituted SiC are shown in Fig.\ref{fig:4HCDOS} and Fig.\ref{fig:4HSiDOS}. In C-sub case, the defect level extends to the conduction band, which should disappear if the band gap issue is fix. In Si-sub case, the defect band is located near the top of the host's valence band. Differently, the defect band is located near the top of the host's valence band in Si-sub case. This is mainly resulted from the charge transfer between silicon and carbon atoms. The coulomb effect from the NN carbons pushes the defect level towards the valence band. The broadened peak of the defect states compared with that in cubic SiC is due to the lowering of symmetry. The bonding mechanism in 4H-SiC is essentially the same as cubic SiC except for the crystal splitting as shown in Fig.\ref{SiCstack}(b).

\begin{figure}[h!]
  \centering
    \includegraphics[width=0.5\textwidth]{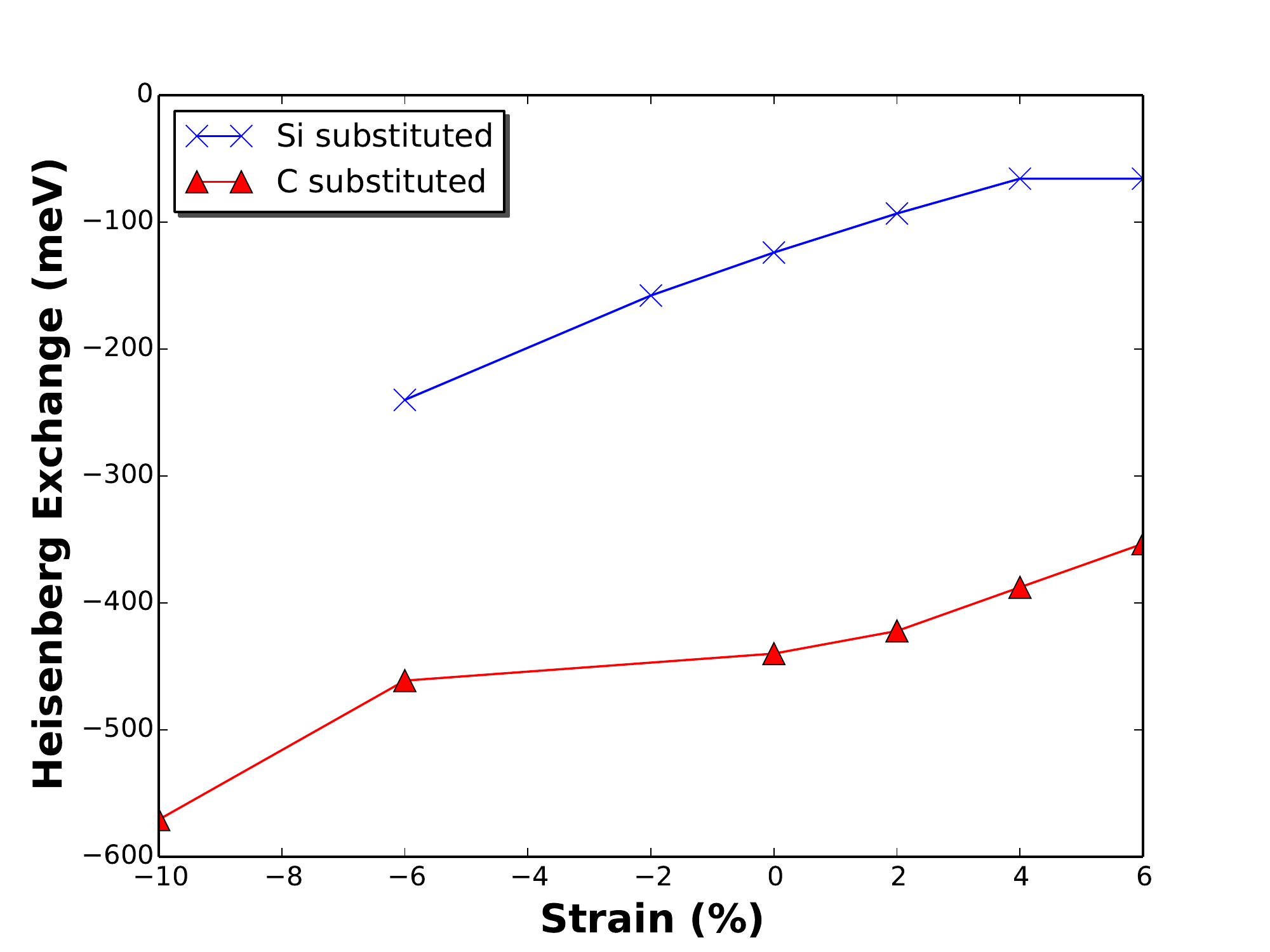}
  \caption{Strain dependence of the Heisenberg exchange energy.}
  \label{formationE4H}
\end{figure}

The strain dependences of Heisenberg exchange energy in C-sub and Si-sub cases are plotted in Fig.\ref{formationE4H}. For both Ni$_\textrm{Si}$ and Ni$_\textrm{C}$, there's no FM-AFM transition happening as we saw in the 3C-SiC. As the strain is raised from -10\% to +6\%, the exchange energy increases linearly with a rising rate of 0.0011 \%/meV. This is mainly due to that the increasing hopping probability as the Ni-NN distance can't overcome the crystal field splitting (CFS) so as to fail triggering the magnetic transition. More specificly, the $t_2$ level is splitted into an $a_1$ and $e$ levels, of which the degeneracies are two and one respectively. The CFS enegy and spin splitting in $a_1$ and $e$ levels possess the relationship that: $E_{CFS}>\frac{1}{2}(E_{ex1}+E_{ex2})$. Therefore, the spin-up and spin-dn $a_1$ levels are fully occupied. The transition will happen only when this inequality is broken.

\subsection{Cr in silicon carbide}
\label{Cr_SiC}
\begin{figure}[h!]
  \centering
    \includegraphics[width=0.5\textwidth]{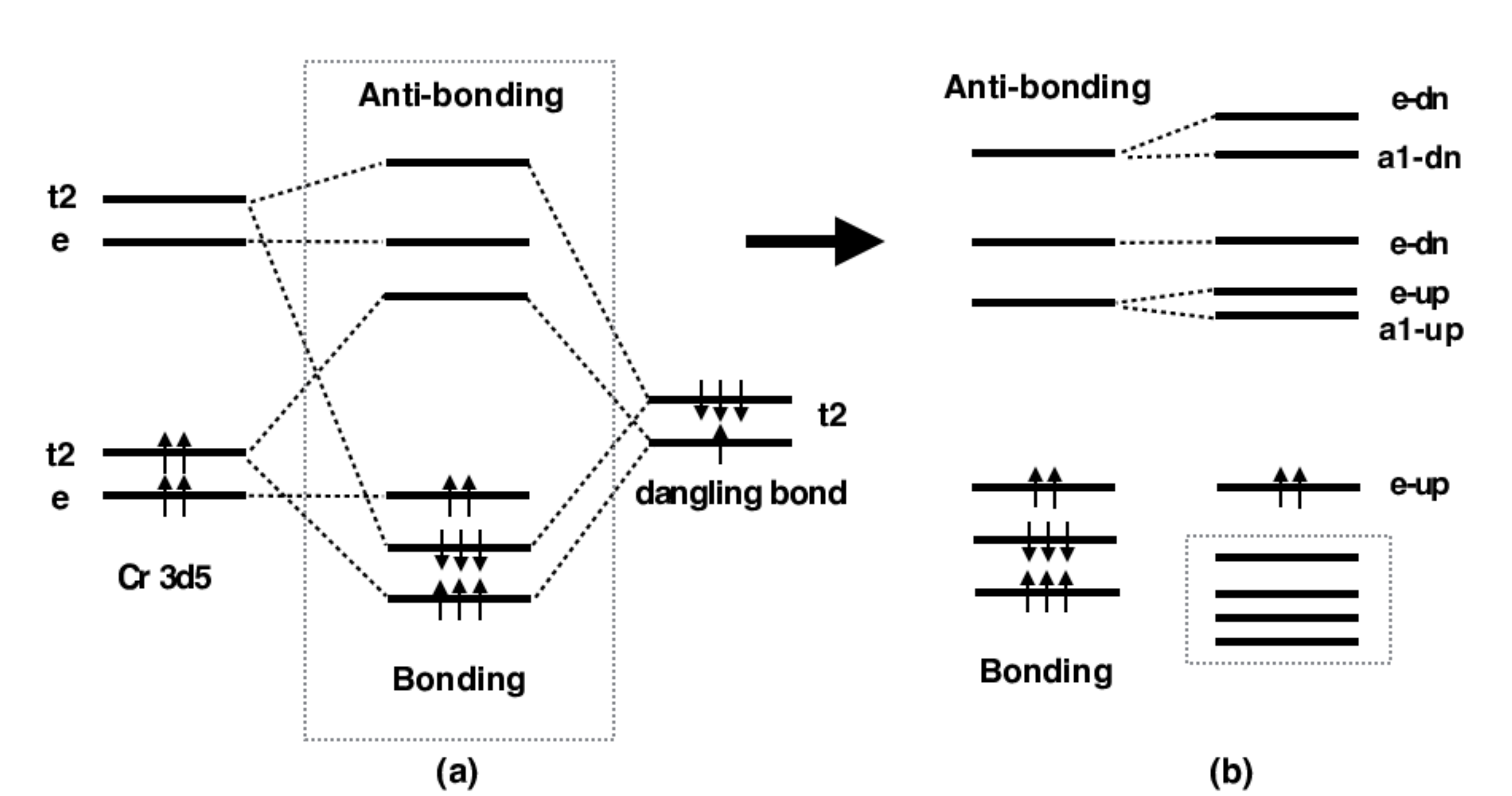}
  \caption{Electron configuration of Cr doped SiC. (a) Bonding mechanism of silicon substituted Cr in 3C-SiC; (b) Level splitting due to symmetry lowering in 4H-SiC.}
\label{Bond_Cr}
\end{figure}

Except for nickel, other transition-metal single dopants is potential to have similar performance in SiC\cite{Chanierprb2012} due to their characters including:
\begin{itemize}
\item the d levels can hybridize with the p orbitals of the dangling bond strongly ($t_2$ levels) and weakly (e levels)\cite{Harold1980};
\item the angular moment 1triplets won't be splitted due to the high symmetry of the crystal field;
\item the larger spin-orbital coupling in d electrons makes high speed electrical field spin manipulation feasible\cite{Tang2006}.
\end{itemize} 
Therefore, it's worth of exploring the combinations of SiC and other transition-metal elements. Starting from 3C-SiC, the magnetism of chromium dopant comes from the highly localized crystal field splitted e level due to four less electrons compared with nickel. By ordering the active electrons from low to high levels, the $t_2$ bonding levels are filled and two additional electrons occupies the e levels as shown in Fig.\ref{Bond_Cr}(a). The stability of Si and C substituted Cr can be evaluated by their formation energy:
\begin{equation}
E^f[X]=E_{tot}[X]-E_p+\mu_{Ni}-\mu_{X},
\end{equation}
where $E_p$ is the total energy of pure SiC, $\mu_\textrm{Ni}$ and $\mu_{X}$ are the chemical potentials of Ni and substituted atom X. The chemical potentials of atoms are chosen to be the values of their bulk materials. Without an external stress, the $E^f$ of Si-sub Cr (1.16 eV) is much lower than that of C-sub Cr (7.72 eV), which indicates that most of the substitutional Cr complexes are formed in the Si sites. 

\begin{figure}
    \begin{subfigure}[b]{0.5\textwidth}
        \includegraphics[width=\textwidth]{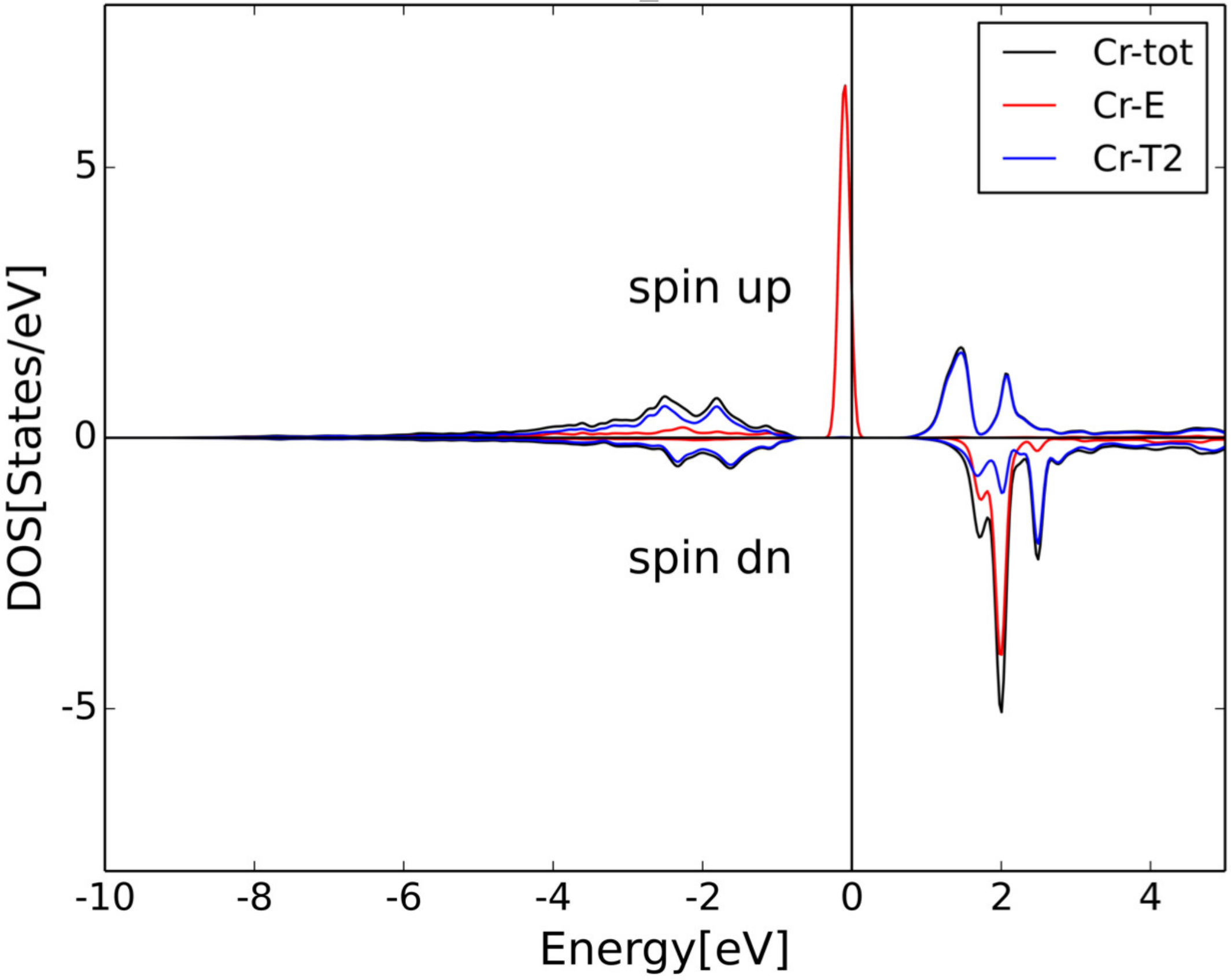}
        \caption{}
        \label{fig:3CSiDOS_Cr}
    \end{subfigure}
    ~ 
    \begin{subfigure}[b]{0.5\textwidth}
        \includegraphics[width=\textwidth]{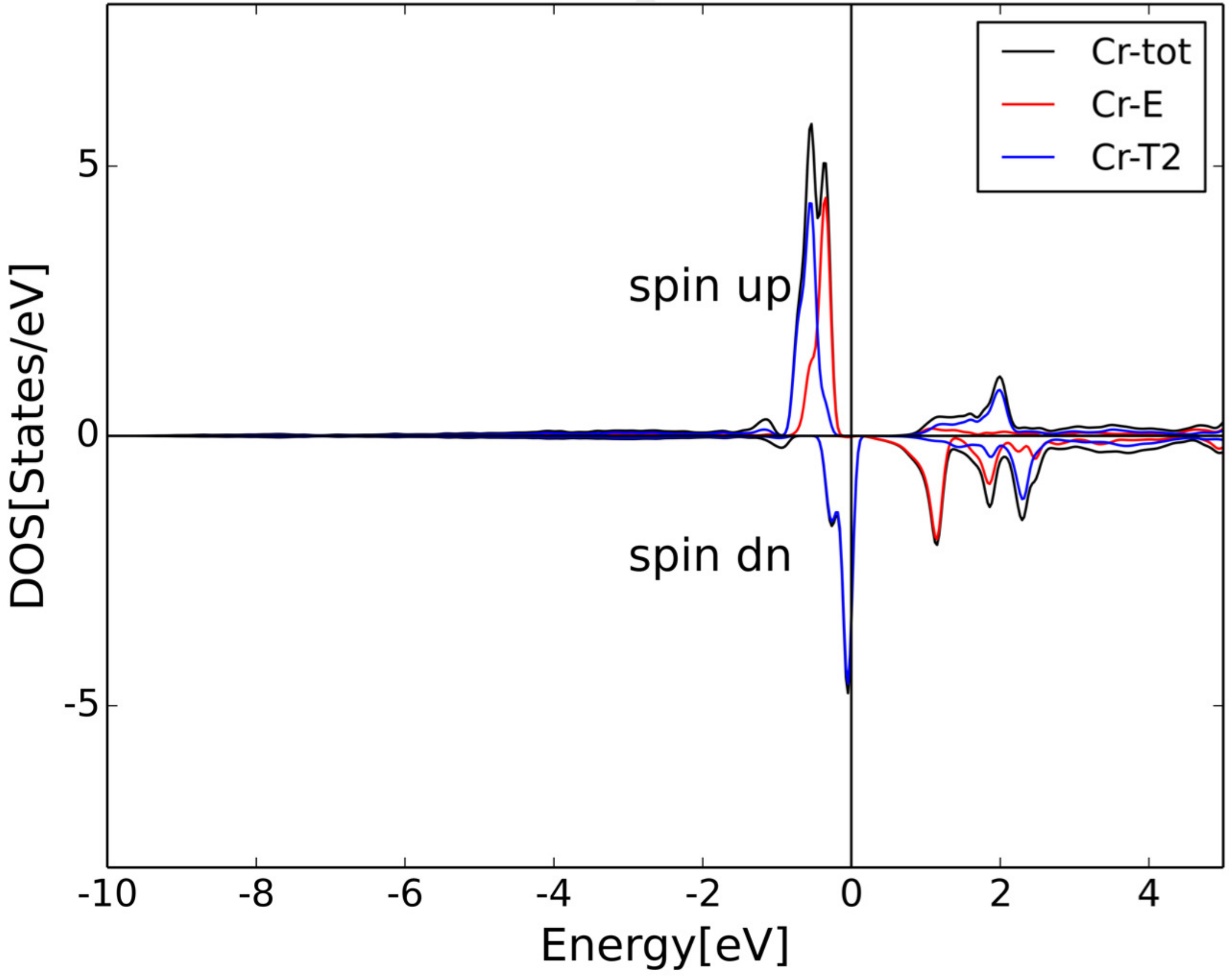}
        \caption{}
        \label{fig:3CDOS_Cr}
    \end{subfigure}
    \caption{DOS of C-sub (a) and Si-sub (b) Ni in 4H-SiC.}
    \label{3CDOS_Cr}
\end{figure}

This bonding mechanism can be verified by looking at the DOS shown in Fig.\ref{3CDOS_Cr}. Here, all the setup is the same as that in nickel calculation. In the C-sub case (Fig.\ref{fig:3CSiDOS_Cr}), the localized $e$ levels are located in the middle of the bandgap coinciding with $t_2$ levels, which is different from Ni dopant. In the Si-sub case (Fig.\ref{fig:3CSiDOS_Cr}), the electronic structure resembles the Ni dopant to a very high extent. The majority e level is located in the middle of bandgap. But the minority e level (2 eV) is pushed into the conduction band due to the much narrower bandgap compared with diamond. The narrow peak of this e level implies it's highly localized. As a result, the spin residing in Cr complex is mainly assigned to the localized e levels. Our calculation also shows that this property makes the spin state insensitive to the applied strain, which holds in both Si and C substituted case. The two unoccupied $t_2$ can be clearly observed at 1.7 eV and 2.5 eV. The spin-up and spin-down $t_2$ states are spread around and hybridize with the $e$ states. The optical spin-flip transition from $e$ up and $e$ down state is feasible due to the spin-orbit interaction, which can mix the $e$ down and $t_2$ states. This transition indicates quite a bit of potential for manipulating the spin system in its excited states

\begin{figure}[h!]
  \centering
    \includegraphics[width=0.5\textwidth]{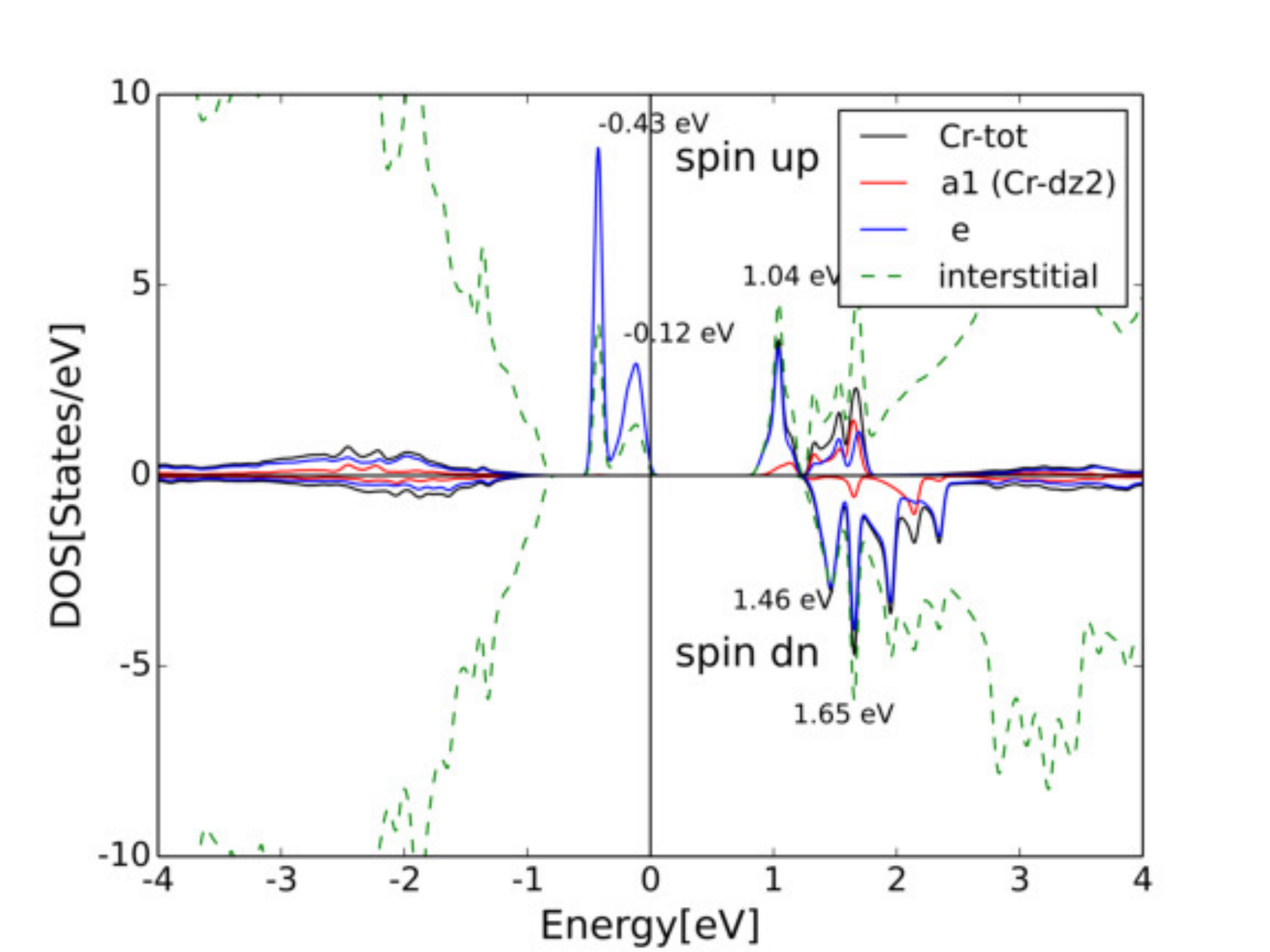}
  \caption{DOS of Silicon substituted Cr in 4H-SiC.}
\label{4HDOS_Cr}
\end{figure}

Compared with the 3C-SiC, the $C_{3v}$ symmetry of 4H-SiC gives rise to a more complicated electronic configuration. As shown in Fig.\ref{Bond_Cr}(b), the $t_2$ level in $T_d$ symmetry is splitted up to a $e$ and down to $a_1$ levels. Again, the detailed electronic configuration can be verified in the DOS shown in Fig.\ref{4HDOS_Cr}. The localized $e$ pair remains unchanged in the bandgap and conduction band. The $t_2$ derived $a_1$ can still bridge these two states through spin-orbit interaction.  



\section{Conclusions}
\label{conclusion}
In this paper, we investigated the strain dependence of magnetic property of single substitutional nickel in 3C and 4H SiC with first principles calculation. For 3C-SiC, we observe FM-AFM transitions around 6\% tensile hydrostatic strain in C-sub nickel and 10\% compressive hydrostatic strain in Si-sub nickel. The Heisenberg exchange-coupling energy is linear to the strain except for a sudden increase at the transition point of C-sub nickel. In contrast, the 4H-SiC doesn't have magnetic phase transition in both substitutional sites but still exhibits a linear relationship between $E_{ex}$ and the strain. Based on the projected orbital occupation through the method of LOMB, we recover the pre-hybridization electron configuration and put it in the $p-d$ hybridization model. We find that a spin pair residing in the substitutional nickel and dangling bond gives rise to the magnetism of the unit cell. By adjusting the distance between the two potential wells, the virtual hopping within the spin pair can be controlled so as to manipulate the spin of the single defect. With an external magnetic field, the qubit can be initialized and manipulated by strain and microwave radiation. Finally, the chromium single dopant in silicon carbide is also investigated. The result shows that a potential d-d spin-flip transition makse the spin manipulation feasible in excited states.

\bibliography{apssamp}

\begin{thebibliography}{33}%
\makeatletter
\providecommand \@ifxundefined [1]{%
 \@ifx{#1\undefined}
}%
\providecommand \@ifnum [1]{%
 \ifnum #1\expandafter \@firstoftwo
 \else \expandafter \@secondoftwo
 \fi
}%
\providecommand \@ifx [1]{%
 \ifx #1\expandafter \@firstoftwo
 \else \expandafter \@secondoftwo
 \fi
}%
\providecommand \natexlab [1]{#1}%
\providecommand \enquote  [1]{``#1''}%
\providecommand \bibnamefont  [1]{#1}%
\providecommand \bibfnamefont [1]{#1}%
\providecommand \citenamefont [1]{#1}%
\providecommand \href@noop [0]{\@secondoftwo}%
\providecommand \href [0]{\begingroup \@sanitize@url \@href}%
\providecommand \@href[1]{\@@startlink{#1}\@@href}%
\providecommand \@@href[1]{\endgroup#1\@@endlink}%
\providecommand \@sanitize@url [0]{\catcode `\\12\catcode `\$12\catcode
  `\&12\catcode `\#12\catcode `\^12\catcode `\_12\catcode `\%12\relax}%
\providecommand \@@startlink[1]{}%
\providecommand \@@endlink[0]{}%
\providecommand \url  [0]{\begingroup\@sanitize@url \@url }%
\providecommand \@url [1]{\endgroup\@href {#1}{\urlprefix }}%
\providecommand \urlprefix  [0]{URL }%
\providecommand \Eprint [0]{\href }%
\providecommand \doibase [0]{http://dx.doi.org/}%
\providecommand \selectlanguage [0]{\@gobble}%
\providecommand \bibinfo  [0]{\@secondoftwo}%
\providecommand \bibfield  [0]{\@secondoftwo}%
\providecommand \translation [1]{[#1]}%
\providecommand \BibitemOpen [0]{}%
\providecommand \bibitemStop [0]{}%
\providecommand \bibitemNoStop [0]{.\EOS\space}%
\providecommand \EOS [0]{\spacefactor3000\relax}%
\providecommand \BibitemShut  [1]{\csname bibitem#1\endcsname}%
\let\auto@bib@innerbib\@empty
\bibitem [{\citenamefont {Isoya}\ \emph {et~al.}(1990)\citenamefont {Isoya},
  \citenamefont {Kanda}, \citenamefont {Norris}, \citenamefont {Tang},\ and\
  \citenamefont {Bowman}}]{Isoya1990}%
  \BibitemOpen
  \bibfield  {author} {\bibinfo {author} {\bibfnamefont {J.}~\bibnamefont
  {Isoya}}, \bibinfo {author} {\bibfnamefont {H.}~\bibnamefont {Kanda}},
  \bibinfo {author} {\bibfnamefont {J.~R.}\ \bibnamefont {Norris}}, \bibinfo
  {author} {\bibfnamefont {J.}~\bibnamefont {Tang}}, \ and\ \bibinfo {author}
  {\bibfnamefont {M.~K.}\ \bibnamefont {Bowman}},\ }\href {\doibase
  10.1103/PhysRevB.41.3905} {\bibfield  {journal} {\bibinfo  {journal} {Phys.
  Rev. B}\ }\textbf {\bibinfo {volume} {41}},\ \bibinfo {pages} {3905}
  (\bibinfo {year} {1990})}\BibitemShut {NoStop}%
\bibitem [{\citenamefont {Iakoubovskii}(2004)}]{Iakoubovskii2004}%
  \BibitemOpen
  \bibfield  {author} {\bibinfo {author} {\bibfnamefont {K.}~\bibnamefont
  {Iakoubovskii}},\ }\href {\doibase 10.1103/PhysRevB.70.205211} {\bibfield
  {journal} {\bibinfo  {journal} {Phys. Rev. B}\ }\textbf {\bibinfo {volume}
  {70}},\ \bibinfo {pages} {205211} (\bibinfo {year} {2004})}\BibitemShut
  {NoStop}%
\bibitem [{\citenamefont {Nadolinny}\ \emph {et~al.}(1999)\citenamefont
  {Nadolinny}, \citenamefont {Yelisseyev}, \citenamefont {Baker}, \citenamefont
  {Newton}, \citenamefont {Twitchen}, \citenamefont {Lawson}, \citenamefont
  {Yuryeva},\ and\ \citenamefont {Feigelson}}]{Nadolinny1999}%
  \BibitemOpen
  \bibfield  {author} {\bibinfo {author} {\bibfnamefont {V.~A.}\ \bibnamefont
  {Nadolinny}}, \bibinfo {author} {\bibfnamefont {A.~P.}\ \bibnamefont
  {Yelisseyev}}, \bibinfo {author} {\bibfnamefont {J.~M.}\ \bibnamefont
  {Baker}}, \bibinfo {author} {\bibfnamefont {M.~E.}\ \bibnamefont {Newton}},
  \bibinfo {author} {\bibfnamefont {D.~J.}\ \bibnamefont {Twitchen}}, \bibinfo
  {author} {\bibfnamefont {S.~C.}\ \bibnamefont {Lawson}}, \bibinfo {author}
  {\bibfnamefont {O.~P.}\ \bibnamefont {Yuryeva}}, \ and\ \bibinfo {author}
  {\bibfnamefont {B.~N.}\ \bibnamefont {Feigelson}},\ }\href
  {http://stacks.iop.org/0953-8984/11/i=38/a=314} {\bibfield  {journal}
  {\bibinfo  {journal} {Journal of Physics: Condensed Matter}\ }\textbf
  {\bibinfo {volume} {11}},\ \bibinfo {pages} {7357} (\bibinfo {year}
  {1999})}\BibitemShut {NoStop}%
\bibitem [{\citenamefont {Gerstmann}\ \emph {et~al.}(2000)\citenamefont
  {Gerstmann}, \citenamefont {Amkreutz},\ and\ \citenamefont
  {Overhof}}]{Gerstmann2000}%
  \BibitemOpen
  \bibfield  {author} {\bibinfo {author} {\bibfnamefont {U.}~\bibnamefont
  {Gerstmann}}, \bibinfo {author} {\bibfnamefont {M.}~\bibnamefont {Amkreutz}},
  \ and\ \bibinfo {author} {\bibfnamefont {H.}~\bibnamefont {Overhof}},\ }\href
  {\doibase 10.1002/(SICI)1521-3951(200001)217:1<665::AID-PSSB665>3.0.CO;2-B}
  {\bibfield  {journal} {\bibinfo  {journal} {physica status solidi (b)}\
  }\textbf {\bibinfo {volume} {217}},\ \bibinfo {pages} {665} (\bibinfo {year}
  {2000})}\BibitemShut {NoStop}%
\bibitem [{\citenamefont {Johnston}\ and\ \citenamefont
  {Mainwood}(2003)}]{Johnston2003}%
  \BibitemOpen
  \bibfield  {author} {\bibinfo {author} {\bibfnamefont {K.}~\bibnamefont
  {Johnston}}\ and\ \bibinfo {author} {\bibfnamefont {A.}~\bibnamefont
  {Mainwood}},\ }\href {\doibase
  http://dx.doi.org/10.1016/S0925-9635(02)00389-8} {\bibfield  {journal}
  {\bibinfo  {journal} {Diamond and Related Materials}\ }\textbf {\bibinfo
  {volume} {12}},\ \bibinfo {pages} {516 } (\bibinfo {year} {2003})},\ \bibinfo
  {note} {13th European Conference on Diamond, Diamond-Like Materials, Carbon
  Nanotubes, Nitrides and Silicon Carbide}\BibitemShut {NoStop}%
\bibitem [{\citenamefont {Pereira}\ \emph {et~al.}(2003)\citenamefont
  {Pereira}, \citenamefont {Gehlhoff}, \citenamefont {Neves},\ and\
  \citenamefont {Sobolev}}]{Pereira2003}%
  \BibitemOpen
  \bibfield  {author} {\bibinfo {author} {\bibfnamefont {R.~N.}\ \bibnamefont
  {Pereira}}, \bibinfo {author} {\bibfnamefont {W.}~\bibnamefont {Gehlhoff}},
  \bibinfo {author} {\bibfnamefont {A.~J.}\ \bibnamefont {Neves}}, \ and\
  \bibinfo {author} {\bibfnamefont {N.~A.}\ \bibnamefont {Sobolev}},\ }\href
  {http://stacks.iop.org/0953-8984/15/i=17/a=305} {\bibfield  {journal}
  {\bibinfo  {journal} {Journal of Physics: Condensed Matter}\ }\textbf
  {\bibinfo {volume} {15}},\ \bibinfo {pages} {2493} (\bibinfo {year}
  {2003})}\BibitemShut {NoStop}%
\bibitem [{\citenamefont {Gaebel}\ \emph {et~al.}(2004)\citenamefont {Gaebel},
  \citenamefont {Popa}, \citenamefont {Gruber}, \citenamefont {Domhan},
  \citenamefont {Jelezko},\ and\ \citenamefont {Wrachtrup}}]{Gaebel2004}%
  \BibitemOpen
  \bibfield  {author} {\bibinfo {author} {\bibfnamefont {T.}~\bibnamefont
  {Gaebel}}, \bibinfo {author} {\bibfnamefont {I.}~\bibnamefont {Popa}},
  \bibinfo {author} {\bibfnamefont {A.}~\bibnamefont {Gruber}}, \bibinfo
  {author} {\bibfnamefont {M.}~\bibnamefont {Domhan}}, \bibinfo {author}
  {\bibfnamefont {F.}~\bibnamefont {Jelezko}}, \ and\ \bibinfo {author}
  {\bibfnamefont {J.}~\bibnamefont {Wrachtrup}},\ }\href
  {http://stacks.iop.org/1367-2630/6/i=1/a=098} {\bibfield  {journal} {\bibinfo
   {journal} {New Journal of Physics}\ }\textbf {\bibinfo {volume} {6}},\
  \bibinfo {pages} {98} (\bibinfo {year} {2004})}\BibitemShut {NoStop}%
\bibitem [{\citenamefont {Rabeau}\ \emph {et~al.}(2005)\citenamefont {Rabeau},
  \citenamefont {Chin}, \citenamefont {Prawer}, \citenamefont {Jelezko},
  \citenamefont {Gaebel},\ and\ \citenamefont {Wrachtrup}}]{Rabeau2005}%
  \BibitemOpen
  \bibfield  {author} {\bibinfo {author} {\bibfnamefont {J.~R.}\ \bibnamefont
  {Rabeau}}, \bibinfo {author} {\bibfnamefont {Y.~L.}\ \bibnamefont {Chin}},
  \bibinfo {author} {\bibfnamefont {S.}~\bibnamefont {Prawer}}, \bibinfo
  {author} {\bibfnamefont {F.}~\bibnamefont {Jelezko}}, \bibinfo {author}
  {\bibfnamefont {T.}~\bibnamefont {Gaebel}}, \ and\ \bibinfo {author}
  {\bibfnamefont {J.}~\bibnamefont {Wrachtrup}},\ }\href {\doibase
  http://dx.doi.org/10.1063/1.1896088} {\bibfield  {journal} {\bibinfo
  {journal} {Applied Physics Letters}\ }\textbf {\bibinfo {volume} {86}},\
  \bibinfo {eid} {131926} (\bibinfo {year} {2005}),\
  http://dx.doi.org/10.1063/1.1896088}\BibitemShut {NoStop}%
\bibitem [{\citenamefont {Wu}\ \emph {et~al.}(2007)\citenamefont {Wu},
  \citenamefont {Rabeau}, \citenamefont {Roger}, \citenamefont {Treussart},
  \citenamefont {Zeng}, \citenamefont {Grangier}, \citenamefont {Prawer},\ and\
  \citenamefont {Roch}}]{Wu2007}%
  \BibitemOpen
  \bibfield  {author} {\bibinfo {author} {\bibfnamefont {E.}~\bibnamefont
  {Wu}}, \bibinfo {author} {\bibfnamefont {J.~R.}\ \bibnamefont {Rabeau}},
  \bibinfo {author} {\bibfnamefont {G.}~\bibnamefont {Roger}}, \bibinfo
  {author} {\bibfnamefont {F.}~\bibnamefont {Treussart}}, \bibinfo {author}
  {\bibfnamefont {H.}~\bibnamefont {Zeng}}, \bibinfo {author} {\bibfnamefont
  {P.}~\bibnamefont {Grangier}}, \bibinfo {author} {\bibfnamefont
  {S.}~\bibnamefont {Prawer}}, \ and\ \bibinfo {author} {\bibfnamefont {J.-F.}\
  \bibnamefont {Roch}},\ }\href {http://stacks.iop.org/1367-2630/9/i=12/a=434}
  {\bibfield  {journal} {\bibinfo  {journal} {New Journal of Physics}\ }\textbf
  {\bibinfo {volume} {9}},\ \bibinfo {pages} {434} (\bibinfo {year}
  {2007})}\BibitemShut {NoStop}%
\bibitem [{\citenamefont {POWELL}\ \emph {et~al.}(2006)\citenamefont {POWELL},
  \citenamefont {JENNY}, \citenamefont {MULLER}, \citenamefont {McD.~HOBGOOD},
  \citenamefont {TSVETKOV}, \citenamefont {LENOARD},\ and\ \citenamefont
  {CARTER}}]{POWELL2006}%
  \BibitemOpen
  \bibfield  {author} {\bibinfo {author} {\bibfnamefont {A.}~\bibnamefont
  {POWELL}}, \bibinfo {author} {\bibfnamefont {J.}~\bibnamefont {JENNY}},
  \bibinfo {author} {\bibfnamefont {S.}~\bibnamefont {MULLER}}, \bibinfo
  {author} {\bibfnamefont {H.}~\bibnamefont {McD.~HOBGOOD}}, \bibinfo {author}
  {\bibfnamefont {V.}~\bibnamefont {TSVETKOV}}, \bibinfo {author}
  {\bibfnamefont {R.}~\bibnamefont {LENOARD}}, \ and\ \bibinfo {author}
  {\bibfnamefont {C.}~\bibnamefont {CARTER}},\ }\href {\doibase
  10.1142/S0129156406004016} {\bibfield  {journal} {\bibinfo  {journal}
  {International Journal of High Speed Electronics and Systems}\ }\textbf
  {\bibinfo {volume} {16}},\ \bibinfo {pages} {751} (\bibinfo {year} {2006})},\
  \Eprint
  {http://arxiv.org/abs/http://www.worldscientific.com/doi/pdf/10.1142/S0129156406004016}
  {http://www.worldscientific.com/doi/pdf/10.1142/S0129156406004016}
  \BibitemShut {NoStop}%
\bibitem [{\citenamefont {Berger}\ \emph {et~al.}(2006)\citenamefont {Berger},
  \citenamefont {Song}, \citenamefont {Li}, \citenamefont {Wu}, \citenamefont
  {Brown}, \citenamefont {Naud}, \citenamefont {Mayou}, \citenamefont {Li},
  \citenamefont {Hass}, \citenamefont {Marchenkov}, \citenamefont {Conrad},
  \citenamefont {First},\ and\ \citenamefont {de~Heer}}]{Berger2006}%
  \BibitemOpen
  \bibfield  {author} {\bibinfo {author} {\bibfnamefont {C.}~\bibnamefont
  {Berger}}, \bibinfo {author} {\bibfnamefont {Z.}~\bibnamefont {Song}},
  \bibinfo {author} {\bibfnamefont {X.}~\bibnamefont {Li}}, \bibinfo {author}
  {\bibfnamefont {X.}~\bibnamefont {Wu}}, \bibinfo {author} {\bibfnamefont
  {N.}~\bibnamefont {Brown}}, \bibinfo {author} {\bibfnamefont
  {C.}~\bibnamefont {Naud}}, \bibinfo {author} {\bibfnamefont {D.}~\bibnamefont
  {Mayou}}, \bibinfo {author} {\bibfnamefont {T.}~\bibnamefont {Li}}, \bibinfo
  {author} {\bibfnamefont {J.}~\bibnamefont {Hass}}, \bibinfo {author}
  {\bibfnamefont {A.~N.}\ \bibnamefont {Marchenkov}}, \bibinfo {author}
  {\bibfnamefont {E.~H.}\ \bibnamefont {Conrad}}, \bibinfo {author}
  {\bibfnamefont {P.~N.}\ \bibnamefont {First}}, \ and\ \bibinfo {author}
  {\bibfnamefont {W.~A.}\ \bibnamefont {de~Heer}},\ }\href {\doibase
  10.1126/science.1125925} {\bibfield  {journal} {\bibinfo  {journal}
  {Science}\ }\textbf {\bibinfo {volume} {312}},\ \bibinfo {pages} {1191}
  (\bibinfo {year} {2006})},\ \Eprint
  {http://arxiv.org/abs/http://science.sciencemag.org/content/312/5777/1191.full.pdf}
  {http://science.sciencemag.org/content/312/5777/1191.full.pdf} \BibitemShut
  {NoStop}%
\bibitem [{\citenamefont {Liu}\ and\ \citenamefont {Edgar}(2002)}]{Liu2002}%
  \BibitemOpen
  \bibfield  {author} {\bibinfo {author} {\bibfnamefont {L.}~\bibnamefont
  {Liu}}\ and\ \bibinfo {author} {\bibfnamefont {J.}~\bibnamefont {Edgar}},\
  }\href {\doibase http://dx.doi.org/10.1016/S0927-796X(02)00008-6} {\bibfield
  {journal} {\bibinfo  {journal} {Materials Science and Engineering: R:
  Reports}\ }\textbf {\bibinfo {volume} {37}},\ \bibinfo {pages} {61 }
  (\bibinfo {year} {2002})}\BibitemShut {NoStop}%
\bibitem [{\citenamefont {Mizuochi}\ \emph {et~al.}(2002)\citenamefont
  {Mizuochi}, \citenamefont {Yamasaki}, \citenamefont {Takizawa}, \citenamefont
  {Morishita}, \citenamefont {Ohshima}, \citenamefont {Itoh},\ and\
  \citenamefont {Isoya}}]{Mizuochi2002}%
  \BibitemOpen
  \bibfield  {author} {\bibinfo {author} {\bibfnamefont {N.}~\bibnamefont
  {Mizuochi}}, \bibinfo {author} {\bibfnamefont {S.}~\bibnamefont {Yamasaki}},
  \bibinfo {author} {\bibfnamefont {H.}~\bibnamefont {Takizawa}}, \bibinfo
  {author} {\bibfnamefont {N.}~\bibnamefont {Morishita}}, \bibinfo {author}
  {\bibfnamefont {T.}~\bibnamefont {Ohshima}}, \bibinfo {author} {\bibfnamefont
  {H.}~\bibnamefont {Itoh}}, \ and\ \bibinfo {author} {\bibfnamefont
  {J.}~\bibnamefont {Isoya}},\ }\href {\doibase 10.1103/PhysRevB.66.235202}
  {\bibfield  {journal} {\bibinfo  {journal} {Phys. Rev. B}\ }\textbf {\bibinfo
  {volume} {66}},\ \bibinfo {pages} {235202} (\bibinfo {year}
  {2002})}\BibitemShut {NoStop}%
\bibitem [{\citenamefont {Son}\ \emph {et~al.}(2003)\citenamefont {Son},
  \citenamefont {Zolnai},\ and\ \citenamefont {Janz\'en}}]{Son2003}%
  \BibitemOpen
  \bibfield  {author} {\bibinfo {author} {\bibfnamefont {N.~T.}\ \bibnamefont
  {Son}}, \bibinfo {author} {\bibfnamefont {Z.}~\bibnamefont {Zolnai}}, \ and\
  \bibinfo {author} {\bibfnamefont {E.}~\bibnamefont {Janz\'en}},\ }\href
  {\doibase 10.1103/PhysRevB.68.205211} {\bibfield  {journal} {\bibinfo
  {journal} {Phys. Rev. B}\ }\textbf {\bibinfo {volume} {68}},\ \bibinfo
  {pages} {205211} (\bibinfo {year} {2003})}\BibitemShut {NoStop}%
\bibitem [{\citenamefont {Son}\ \emph {et~al.}(2006)\citenamefont {Son},
  \citenamefont {Carlsson}, \citenamefont {ul~Hassan}, \citenamefont
  {Janz\'en}, \citenamefont {Umeda}, \citenamefont {Isoya}, \citenamefont
  {Gali}, \citenamefont {Bockstedte}, \citenamefont {Morishita}, \citenamefont
  {Ohshima},\ and\ \citenamefont {Itoh}}]{Son2006}%
  \BibitemOpen
  \bibfield  {author} {\bibinfo {author} {\bibfnamefont {N.~T.}\ \bibnamefont
  {Son}}, \bibinfo {author} {\bibfnamefont {P.}~\bibnamefont {Carlsson}},
  \bibinfo {author} {\bibfnamefont {J.}~\bibnamefont {ul~Hassan}}, \bibinfo
  {author} {\bibfnamefont {E.}~\bibnamefont {Janz\'en}}, \bibinfo {author}
  {\bibfnamefont {T.}~\bibnamefont {Umeda}}, \bibinfo {author} {\bibfnamefont
  {J.}~\bibnamefont {Isoya}}, \bibinfo {author} {\bibfnamefont
  {A.}~\bibnamefont {Gali}}, \bibinfo {author} {\bibfnamefont {M.}~\bibnamefont
  {Bockstedte}}, \bibinfo {author} {\bibfnamefont {N.}~\bibnamefont
  {Morishita}}, \bibinfo {author} {\bibfnamefont {T.}~\bibnamefont {Ohshima}},
  \ and\ \bibinfo {author} {\bibfnamefont {H.}~\bibnamefont {Itoh}},\ }\href
  {\doibase 10.1103/PhysRevLett.96.055501} {\bibfield  {journal} {\bibinfo
  {journal} {Phys. Rev. Lett.}\ }\textbf {\bibinfo {volume} {96}},\ \bibinfo
  {pages} {055501} (\bibinfo {year} {2006})}\BibitemShut {NoStop}%
\bibitem [{\citenamefont {Gali}(2011)}]{GaliBSP2011}%
  \BibitemOpen
  \bibfield  {author} {\bibinfo {author} {\bibfnamefont {A.}~\bibnamefont
  {Gali}},\ }\href {\doibase 10.1002/pssb.201046254} {\bibfield  {journal}
  {\bibinfo  {journal} {physica status solidi (b)}\ }\textbf {\bibinfo {volume}
  {248}},\ \bibinfo {pages} {1337} (\bibinfo {year} {2011})}\BibitemShut
  {NoStop}%
\bibitem [{\citenamefont {Falk}\ \emph {et~al.}(2013)\citenamefont {Falk},
  \citenamefont {Buckley}, \citenamefont {Calusine}, \citenamefont {Koehl},
  \citenamefont {Dobrovitski}, \citenamefont {Politi}, \citenamefont {Zorman},
  \citenamefont {Feng},\ and\ \citenamefont {Awschalom}}]{Falk2013}%
  \BibitemOpen
  \bibfield  {author} {\bibinfo {author} {\bibfnamefont {A.~L.}\ \bibnamefont
  {Falk}}, \bibinfo {author} {\bibfnamefont {B.~B.}\ \bibnamefont {Buckley}},
  \bibinfo {author} {\bibfnamefont {G.}~\bibnamefont {Calusine}}, \bibinfo
  {author} {\bibfnamefont {W.~F.}\ \bibnamefont {Koehl}}, \bibinfo {author}
  {\bibfnamefont {V.~V.}\ \bibnamefont {Dobrovitski}}, \bibinfo {author}
  {\bibfnamefont {A.}~\bibnamefont {Politi}}, \bibinfo {author} {\bibfnamefont
  {C.~A.}\ \bibnamefont {Zorman}}, \bibinfo {author} {\bibfnamefont {P.~X.-L.}\
  \bibnamefont {Feng}}, \ and\ \bibinfo {author} {\bibfnamefont {D.~D.}\
  \bibnamefont {Awschalom}},\ }\href {http://dx.doi.org/10.1038/ncomms2854
  10.1038/ncomms2854} {\bibfield  {journal} {\bibinfo  {journal} {Nat Commun}\
  }\textbf {\bibinfo {volume} {4}},\ \bibinfo {pages} {1819} (\bibinfo {year}
  {2013})}\BibitemShut {NoStop}%
\bibitem [{\citenamefont {Schwarz}\ \emph {et~al.}(2002)\citenamefont
  {Schwarz}, \citenamefont {Blaha},\ and\ \citenamefont
  {Madsen}}]{Schwarz2002}%
  \BibitemOpen
  \bibfield  {author} {\bibinfo {author} {\bibfnamefont {K.}~\bibnamefont
  {Schwarz}}, \bibinfo {author} {\bibfnamefont {P.}~\bibnamefont {Blaha}}, \
  and\ \bibinfo {author} {\bibfnamefont {G.}~\bibnamefont {Madsen}},\ }\href
  {\doibase http://dx.doi.org/10.1016/S0010-4655(02)00206-0} {\bibfield
  {journal} {\bibinfo  {journal} {Computer Physics Communications}\ }\textbf
  {\bibinfo {volume} {147}},\ \bibinfo {pages} {71 } (\bibinfo {year}
  {2002})}\BibitemShut {NoStop}%
\bibitem [{\citenamefont {Perdew}\ \emph {et~al.}(1996)\citenamefont {Perdew},
  \citenamefont {Burke},\ and\ \citenamefont {Ernzerhof}}]{PBE1996}%
  \BibitemOpen
  \bibfield  {author} {\bibinfo {author} {\bibfnamefont {J.~P.}\ \bibnamefont
  {Perdew}}, \bibinfo {author} {\bibfnamefont {K.}~\bibnamefont {Burke}}, \
  and\ \bibinfo {author} {\bibfnamefont {M.}~\bibnamefont {Ernzerhof}},\ }\href
  {\doibase 10.1103/PhysRevLett.77.3865} {\bibfield  {journal} {\bibinfo
  {journal} {Phys. Rev. Lett.}\ }\textbf {\bibinfo {volume} {77}},\ \bibinfo
  {pages} {3865} (\bibinfo {year} {1996})}\BibitemShut {NoStop}%
\bibitem [{\citenamefont {Monkhorst}\ and\ \citenamefont
  {Pack}(1976)}]{Monkhorst1976}%
  \BibitemOpen
  \bibfield  {author} {\bibinfo {author} {\bibfnamefont {H.~J.}\ \bibnamefont
  {Monkhorst}}\ and\ \bibinfo {author} {\bibfnamefont {J.~D.}\ \bibnamefont
  {Pack}},\ }\href {\doibase 10.1103/PhysRevB.13.5188} {\bibfield  {journal}
  {\bibinfo  {journal} {Phys. Rev. B}\ }\textbf {\bibinfo {volume} {13}},\
  \bibinfo {pages} {5188} (\bibinfo {year} {1976})}\BibitemShut {NoStop}%
\bibitem [{\citenamefont {Chanier}\ \emph
  {et~al.}(2012{\natexlab{a}})\citenamefont {Chanier}, \citenamefont {Pryor},\
  and\ \citenamefont {Flatté}}]{Chanier2012}%
  \BibitemOpen
  \bibfield  {author} {\bibinfo {author} {\bibfnamefont {T.}~\bibnamefont
  {Chanier}}, \bibinfo {author} {\bibfnamefont {C.~E.}\ \bibnamefont {Pryor}},
  \ and\ \bibinfo {author} {\bibfnamefont {M.~E.}\ \bibnamefont {Flatté}},\
  }\href {http://stacks.iop.org/0295-5075/99/i=6/a=67006} {\bibfield  {journal}
  {\bibinfo  {journal} {EPL (Europhysics Letters)}\ }\textbf {\bibinfo {volume}
  {99}},\ \bibinfo {pages} {67006} (\bibinfo {year}
  {2012}{\natexlab{a}})}\BibitemShut {NoStop}%
\bibitem [{\citenamefont {Koepernik}\ and\ \citenamefont
  {Eschrig}(1999)}]{Koepernik1999}%
  \BibitemOpen
  \bibfield  {author} {\bibinfo {author} {\bibfnamefont {K.}~\bibnamefont
  {Koepernik}}\ and\ \bibinfo {author} {\bibfnamefont {H.}~\bibnamefont
  {Eschrig}},\ }\href {\doibase 10.1103/PhysRevB.59.1743} {\bibfield  {journal}
  {\bibinfo  {journal} {Phys. Rev. B}\ }\textbf {\bibinfo {volume} {59}},\
  \bibinfo {pages} {1743} (\bibinfo {year} {1999})}\BibitemShut {NoStop}%
\bibitem [{\citenamefont {Zhao}\ and\ \citenamefont
  {Bagayoko}(2000)}]{Zhao2000}%
  \BibitemOpen
  \bibfield  {author} {\bibinfo {author} {\bibfnamefont {G.~L.}\ \bibnamefont
  {Zhao}}\ and\ \bibinfo {author} {\bibfnamefont {D.}~\bibnamefont
  {Bagayoko}},\ }\href {http://stacks.iop.org/1367-2630/2/i=1/a=316} {\bibfield
   {journal} {\bibinfo  {journal} {New Journal of Physics}\ }\textbf {\bibinfo
  {volume} {2}},\ \bibinfo {pages} {16} (\bibinfo {year} {2000})}\BibitemShut
  {NoStop}%
\bibitem [{\citenamefont {Ferr\'on}\ \emph {et~al.}(2015)\citenamefont
  {Ferr\'on}, \citenamefont {Lado},\ and\ \citenamefont
  {Fern\'andez-Rossier}}]{PhysRevB.92.174407}%
  \BibitemOpen
  \bibfield  {author} {\bibinfo {author} {\bibfnamefont {A.}~\bibnamefont
  {Ferr\'on}}, \bibinfo {author} {\bibfnamefont {J.~L.}\ \bibnamefont {Lado}},
  \ and\ \bibinfo {author} {\bibfnamefont {J.}~\bibnamefont
  {Fern\'andez-Rossier}},\ }\href {\doibase 10.1103/PhysRevB.92.174407}
  {\bibfield  {journal} {\bibinfo  {journal} {Phys. Rev. B}\ }\textbf {\bibinfo
  {volume} {92}},\ \bibinfo {pages} {174407} (\bibinfo {year}
  {2015})}\BibitemShut {NoStop}%
\bibitem [{\citenamefont {Los}\ \emph {et~al.}(2007)\citenamefont {Los},
  \citenamefont {Timoshevskii}, \citenamefont {Los},\ and\ \citenamefont
  {Kalkuta}}]{Los2007}%
  \BibitemOpen
  \bibfield  {author} {\bibinfo {author} {\bibfnamefont {A.~V.}\ \bibnamefont
  {Los}}, \bibinfo {author} {\bibfnamefont {A.~N.}\ \bibnamefont
  {Timoshevskii}}, \bibinfo {author} {\bibfnamefont {V.~F.}\ \bibnamefont
  {Los}}, \ and\ \bibinfo {author} {\bibfnamefont {S.~A.}\ \bibnamefont
  {Kalkuta}},\ }\href {\doibase 10.1103/PhysRevB.76.165204} {\bibfield
  {journal} {\bibinfo  {journal} {Phys. Rev. B}\ }\textbf {\bibinfo {volume}
  {76}},\ \bibinfo {pages} {165204} (\bibinfo {year} {2007})}\BibitemShut
  {NoStop}%
\bibitem [{\citenamefont {Mott}(1969)}]{Mott1968}%
  \BibitemOpen
  \bibfield  {author} {\bibinfo {author} {\bibfnamefont {N.~F.}\ \bibnamefont
  {Mott}},\ }\href {\doibase 10.1080/14786436908216338} {\bibfield  {journal}
  {\bibinfo  {journal} {Philosophical Magazine}\ }\textbf {\bibinfo {volume}
  {19}},\ \bibinfo {pages} {835} (\bibinfo {year} {1969})},\ \Eprint
  {http://arxiv.org/abs/http://dx.doi.org/10.1080/14786436908216338}
  {http://dx.doi.org/10.1080/14786436908216338} \BibitemShut {NoStop}%
\bibitem [{\citenamefont {Kunc}\ \emph {et~al.}(1975)\citenamefont {Kunc},
  \citenamefont {Balkanski},\ and\ \citenamefont {Nusimovici}}]{stiffness}%
  \BibitemOpen
  \bibfield  {author} {\bibinfo {author} {\bibfnamefont {K.}~\bibnamefont
  {Kunc}}, \bibinfo {author} {\bibfnamefont {M.}~\bibnamefont {Balkanski}}, \
  and\ \bibinfo {author} {\bibfnamefont {M.~A.}\ \bibnamefont {Nusimovici}},\
  }\href {\doibase 10.1002/pssb.2220720125} {\bibfield  {journal} {\bibinfo
  {journal} {physica status solidi (b)}\ }\textbf {\bibinfo {volume} {72}},\
  \bibinfo {pages} {229} (\bibinfo {year} {1975})}\BibitemShut {NoStop}%
\bibitem [{\citenamefont {Dubrovinsky}\ \emph {et~al.}(2015)\citenamefont
  {Dubrovinsky}, \citenamefont {Dubrovinskaia}, \citenamefont {Bykova},
  \citenamefont {Bykov}, \citenamefont {Prakapenka}, \citenamefont {Prescher},
  \citenamefont {Glazyrin}, \citenamefont {Liermann}, \citenamefont {Hanfland},
  \citenamefont {Ekholm}, \citenamefont {Feng}, \citenamefont {Pourovskii},
  \citenamefont {Katsnelson}, \citenamefont {Wills},\ and\ \citenamefont
  {Abrikosov}}]{Dubrovinsky2015}%
  \BibitemOpen
  \bibfield  {author} {\bibinfo {author} {\bibfnamefont {L.}~\bibnamefont
  {Dubrovinsky}}, \bibinfo {author} {\bibfnamefont {N.}~\bibnamefont
  {Dubrovinskaia}}, \bibinfo {author} {\bibfnamefont {E.}~\bibnamefont
  {Bykova}}, \bibinfo {author} {\bibfnamefont {M.}~\bibnamefont {Bykov}},
  \bibinfo {author} {\bibfnamefont {V.}~\bibnamefont {Prakapenka}}, \bibinfo
  {author} {\bibfnamefont {C.}~\bibnamefont {Prescher}}, \bibinfo {author}
  {\bibfnamefont {K.}~\bibnamefont {Glazyrin}}, \bibinfo {author}
  {\bibfnamefont {H.-P.}\ \bibnamefont {Liermann}}, \bibinfo {author}
  {\bibfnamefont {M.}~\bibnamefont {Hanfland}}, \bibinfo {author}
  {\bibfnamefont {M.}~\bibnamefont {Ekholm}}, \bibinfo {author} {\bibfnamefont
  {Q.}~\bibnamefont {Feng}}, \bibinfo {author} {\bibfnamefont {L.~V.}\
  \bibnamefont {Pourovskii}}, \bibinfo {author} {\bibfnamefont {M.~I.}\
  \bibnamefont {Katsnelson}}, \bibinfo {author} {\bibfnamefont {J.~M.}\
  \bibnamefont {Wills}}, \ and\ \bibinfo {author} {\bibfnamefont {I.~A.}\
  \bibnamefont {Abrikosov}},\ }\href {http://dx.doi.org/10.1038/nature14681
  10.1038/nature14681} {\bibfield  {journal} {\bibinfo  {journal} {Nature}\
  }\textbf {\bibinfo {volume} {525}},\ \bibinfo {pages} {226} (\bibinfo {year}
  {2015})}\BibitemShut {NoStop}%
\bibitem [{\citenamefont {Miao}\ and\ \citenamefont
  {Lambrecht}(2006)}]{Miao2006}%
  \BibitemOpen
  \bibfield  {author} {\bibinfo {author} {\bibfnamefont {M.~S.}\ \bibnamefont
  {Miao}}\ and\ \bibinfo {author} {\bibfnamefont {W.~R.~L.}\ \bibnamefont
  {Lambrecht}},\ }\href {\doibase 10.1103/PhysRevB.74.235218} {\bibfield
  {journal} {\bibinfo  {journal} {Phys. Rev. B}\ }\textbf {\bibinfo {volume}
  {74}},\ \bibinfo {pages} {235218} (\bibinfo {year} {2006})}\BibitemShut
  {NoStop}%
\bibitem [{\citenamefont {w.~Anderson}\ \emph {et~al.}(1972)\citenamefont
  {w.~Anderson}, \citenamefont {Halperin},\ and\ \citenamefont
  {c.~M.~Varma}}]{Anderson1972}%
  \BibitemOpen
  \bibfield  {author} {\bibinfo {author} {\bibfnamefont {P.}~\bibnamefont
  {w.~Anderson}}, \bibinfo {author} {\bibfnamefont {B.~I.}\ \bibnamefont
  {Halperin}}, \ and\ \bibinfo {author} {\bibnamefont {c.~M.~Varma}},\ }\href
  {\doibase 10.1080/14786437208229210} {\bibfield  {journal} {\bibinfo
  {journal} {Philosophical Magazine}\ }\textbf {\bibinfo {volume} {25}},\
  \bibinfo {pages} {1} (\bibinfo {year} {1972})},\ \Eprint
  {http://arxiv.org/abs/http://dx.doi.org/10.1080/14786437208229210}
  {http://dx.doi.org/10.1080/14786437208229210} \BibitemShut {NoStop}%
\bibitem [{\citenamefont {Chanier}\ \emph
  {et~al.}(2012{\natexlab{b}})\citenamefont {Chanier}, \citenamefont {Pryor},\
  and\ \citenamefont {Flatt\'e}}]{Chanierprb2012}%
  \BibitemOpen
  \bibfield  {author} {\bibinfo {author} {\bibfnamefont {T.}~\bibnamefont
  {Chanier}}, \bibinfo {author} {\bibfnamefont {C.}~\bibnamefont {Pryor}}, \
  and\ \bibinfo {author} {\bibfnamefont {M.~E.}\ \bibnamefont {Flatt\'e}},\
  }\href {\doibase 10.1103/PhysRevB.86.085203} {\bibfield  {journal} {\bibinfo
  {journal} {Phys. Rev. B}\ }\textbf {\bibinfo {volume} {86}},\ \bibinfo
  {pages} {085203} (\bibinfo {year} {2012}{\natexlab{b}})}\BibitemShut
  {NoStop}%
\bibitem [{\citenamefont {Hjalmarson}\ \emph {et~al.}(1980)\citenamefont
  {Hjalmarson}, \citenamefont {Vogl}, \citenamefont {Wolford},\ and\
  \citenamefont {Dow}}]{Harold1980}%
  \BibitemOpen
  \bibfield  {author} {\bibinfo {author} {\bibfnamefont {H.~P.}\ \bibnamefont
  {Hjalmarson}}, \bibinfo {author} {\bibfnamefont {P.}~\bibnamefont {Vogl}},
  \bibinfo {author} {\bibfnamefont {D.~J.}\ \bibnamefont {Wolford}}, \ and\
  \bibinfo {author} {\bibfnamefont {J.~D.}\ \bibnamefont {Dow}},\ }\href
  {\doibase 10.1103/PhysRevLett.44.810} {\bibfield  {journal} {\bibinfo
  {journal} {Phys. Rev. Lett.}\ }\textbf {\bibinfo {volume} {44}},\ \bibinfo
  {pages} {810} (\bibinfo {year} {1980})}\BibitemShut {NoStop}%
\bibitem [{\citenamefont {Tang}\ \emph {et~al.}(2006)\citenamefont {Tang},
  \citenamefont {Levy},\ and\ \citenamefont {Flatt\'e}}]{Tang2006}%
  \BibitemOpen
  \bibfield  {author} {\bibinfo {author} {\bibfnamefont {J.-M.}\ \bibnamefont
  {Tang}}, \bibinfo {author} {\bibfnamefont {J.}~\bibnamefont {Levy}}, \ and\
  \bibinfo {author} {\bibfnamefont {M.~E.}\ \bibnamefont {Flatt\'e}},\ }\href
  {\doibase 10.1103/PhysRevLett.97.106803} {\bibfield  {journal} {\bibinfo
  {journal} {Phys. Rev. Lett.}\ }\textbf {\bibinfo {volume} {97}},\ \bibinfo
  {pages} {106803} (\bibinfo {year} {2006})}\BibitemShut {NoStop}%
\end{thebibliography}%

\end{document}